\documentclass[prb, floatfix, nobibnotes, reprint, superscriptaddress]{revtex4-1}
\usepackage{graphicx} 
\usepackage{xcolor}
\usepackage{multirow}
\usepackage[none]{hyphenat}
\usepackage{ragged2e}
\usepackage{booktabs} 
\usepackage{eqnarray, amsmath,blkarray,bigstrut}
\usepackage{mathtools}
\usepackage{url, hyperref}

\usepackage[version=4]{mhchem}

\begin{document}
\title{\papertitle}
\title{Combining Hammett $\sigma$ constants for $\Delta$-machine learning and catalyst discovery}

\author{V. Diana Rakotonirina}
\affiliation{Department of Materials Science and Engineering, University of Toronto, St. George Campus, Toronto, ON, Canada}
\author{Marco Bragato}
\affiliation{Faculty of Physics, University of Vienna, Kolingasse 1416, AT1090 Wien, Austria}
\author{Stefan Heinen}
\affiliation{Vector Institute for Artificial Intelligence, Toronto, ON, M5S 1M1, Canada}
\author{O. Anatole von Lilienfeld}
\email{anatole.vonlilienfeld@utoronto.ca}
\affiliation{Department of Materials Science and Engineering, University of Toronto, St. George Campus, Toronto, ON, Canada}
\affiliation{Vector Institute for Artificial Intelligence, Toronto, ON, M5S 1M1, Canada}
\affiliation{Chemical Physics Theory Group, Department of Chemistry, University of Toronto, St. George Campus, Toronto, ON, Canada}
\affiliation{ML Group, Technische Universit\"at Berlin and Institute for the Foundations of Learning and Data, 10587 Berlin, Germany}
\affiliation{Berlin Institute for the Foundations of Learning and Data, 10587 Berlin, Germany}
\affiliation{Department of Physics, University of Toronto, St. George Campus, Toronto, ON, Canada}
\affiliation{Acceleration Consortium, University of Toronto, Toronto, ON}

\begin{abstract}

We study the applicability of the Hammett-inspired product (HIP) Ansatz to model relative substrate binding within homogenous organometallic catalysis, assigning $\sigma$ and $\rho$ to ligands and metals, respectively. 
Implementing an additive combination (c) rule for obtaining $\sigma$ constants for any ligand pair combination results in a cHIP model that enhances data efficiency in computational ligand tuning.
We show its usefulness (i) as a baseline for $\Delta$-machine learning (ML), and (ii) to identify novel catalyst candidates via volcano plots. 
After testing the combination rule on Hammett constants previously published in the literature, we have generated numerical evidence for the Suzuki-Miyaura (SM) C-C cross-coupling reaction using two synthetic datasets of metallic catalysts (including (10) and (11)-metals Ni, Pd, Pt, and Cu, Ag, Au as well as 96 ligands such as N-heterocyclic carbenes, phosphines, or pyridines). 
When used as a baseline, $\Delta$-ML prediction errors of relative binding decrease systematically with training set size and reach chemical accuracy ($\sim$1 kcal/mol) for 20k training instances. 
Employing the individual ligand constants obtained from cHIP, we report relative substrate binding for a novel dataset consisting of 720 catalysts (not part of training data), of which 145 fall into the most promising range on the volcano plot accounting for oxidative addition, transmetalation, and reductive elimination steps. 
Multiple Ni-based catalysts, e.g. Aphos-Ni-P($t$-Bu)$_3$, are included among these promising candidates, potentially offering dramatic cost savings in experimental applications. 


\end{abstract}

\maketitle

\section{Introduction}
\begin{figure*}
    \centering
   \includegraphics[scale=0.38]{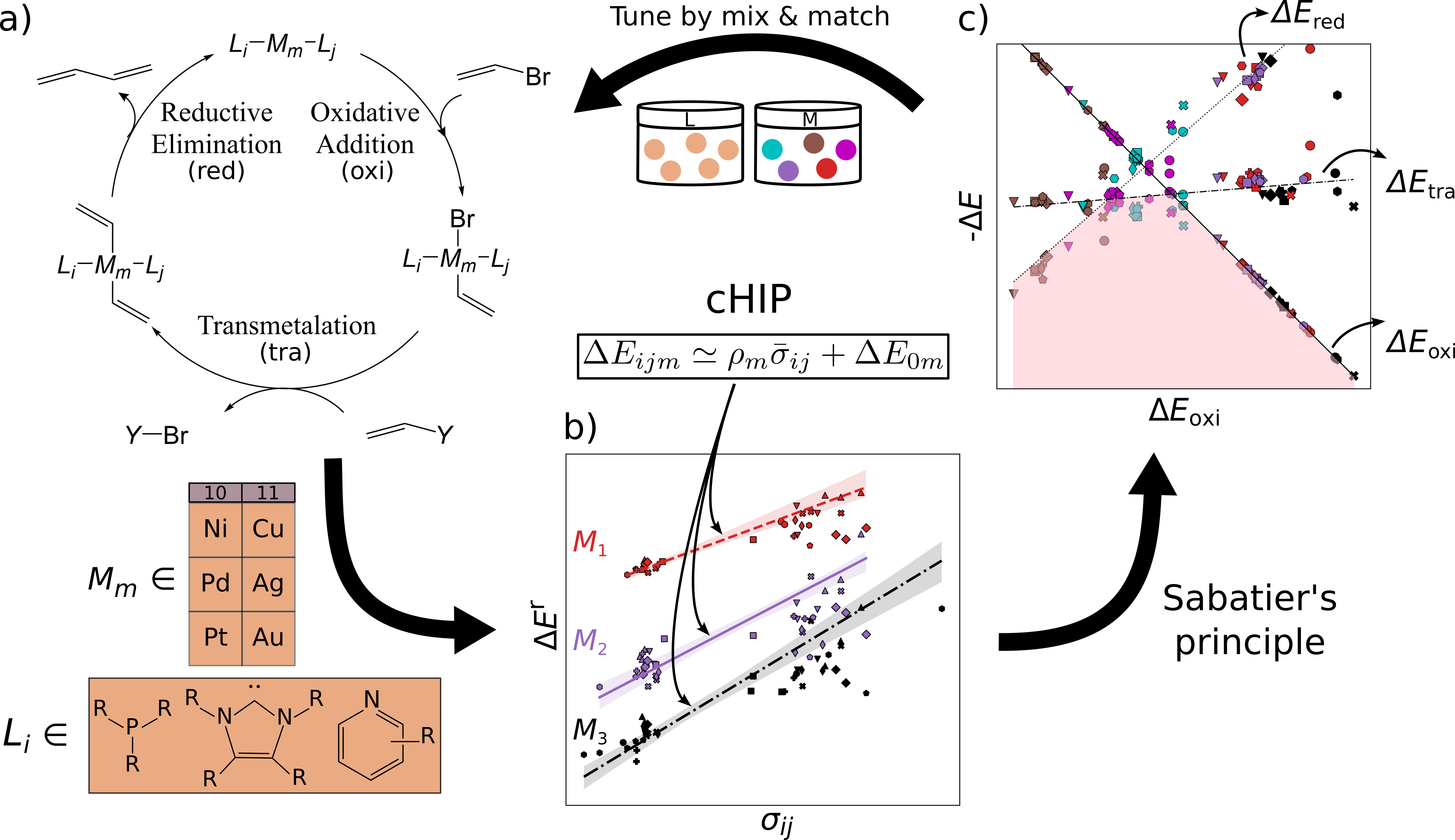}
    \caption{\justifying Vision of iterative catalyst discovery. a) Catalytic cycle of Suzuki-Miyaura C-C cross-coupling for 1,3-butadiene formation with organometallic catalyst $L_{i}-M_m-L_{j}$ and coupling partner $Y$ (\ce{[B(OH)_{2}(O^{t}Bu)]^{-}}) for metals $M$ and exemplary ligands $L$ indicated below. b) Parameters for combination Hammett Inspired Product (cHIP) model of relative binding energies $\Delta E^\mathrm{r}$ are fitted for each complex: $\rho$ for the metal, and averaged $\sigma$ for ligand combinations. c) Volcano plot generated by predicting the relative binding energies corresponding to all three intermediate steps.}
    \label{fig:workflow}
\end{figure*}


A combinatorial approach is crucial for efficiently exploring the vast chemical compound space, facilitating the discovery of new catalysts \cite{pradal2014combinatorial, reetz2008combinatorial}, materials \cite{moulin2012dynamic}, and drugs \cite{liu2017combinatorial} through high-throughput technologies and systematic variation of components \cite{collins2014contemporary, shimizu1998high, isbrandt2019high}. Despite the existence of programs for fragment screening and combinatorial library design \cite{shepherd2014fragment, li2016rapid}, these applications remain predominantly experimental, limiting the ability to freely sample the chemical space.

On the computational front, traditional methods require separate calculations for each complete compound, which is inefficient for large-scale exploration. 
Efforts to reduce the cost of discovering new compounds led to the rapid growth of machine learning (ML) \cite{nandy2021from, kitchin2018machine}, and the combination of experimental and computational techniques via self-driving labs \cite{burger2020sdl, huang2023central} are emerging.
The rise of ML has revolutionized this field by significantly reducing the computational cost of predicting compound properties compared to \textit{ab initio} methods such as density functional theory (DFT) , and as pointed out in Ref. \cite{strieth2020machine,rupp2012coulomb}. 
In homogeneous catalysis, ML techniques such as random forest and linear regression have been used to predict catalyst reactivity \cite{ahneman2018predicting, reid2019holistic, jorner2021organic,dos2021navigating}, while kernel ridge regression (KRR) and neural networks have successfully modeled binding energies for the Suzuki-Miyaura (SM) cross-coupling reaction \cite{meyer2018, schilter2023designing}, relevant in drug synthesis \cite{miyaura1979new, miyaura1995palladium, suzuki2011cross}. 
Additionally, descriptors, or representations, capturing parameters essential in inferring a system's properties have been extensively investigated for different models \cite{rupp2012coulomb, SLATM, BoB, khan2023kernel, damewood2023representations, chang2019hammett}.
Despite the advances, these models often require extensive computations for each catalyst, highlighting the need for a combinatorial strategy that can efficiently explore the catalyst space by integrating the contributions of various building blocks, such as ligands and metals, to optimize performance \cite{han2013Pd_to_Ni, martin2008palladium}.


Linear free energy relationships, such as the Hammett equation \cite{hammett1935some, hammett1937effect}, can be harnessed to build models able to partition systems into fragments, thus eliminating the combinatorial complexity.
Introduced in 1935 \cite{hammett1935some, hammett1937effect}, it has been recognized as a simple but accurate tool for separating substituent and reaction effects on free energy changes.
It was first proposed for benzoic acid derivatives and was subsequently shown to be generalizable. 
Applications to heterocyclic compounds \cite{jaffe1964heterocyclic}, metal-ligand complexes \cite{Chattopadhyay_helper_1976}, and structure-reactivity relationships in cross-coupling reactions \cite{dong2010negishi_hammett} have been reported.
Improvements to better account for steric effects also exist, such as the Taft or the Charton equations \cite{taft1953linear, taft1952polar, taft1952linear, charton1975steric, charton1975steric2, charton1976steric}.
The use of these established parameters is however hindered by the unavailability of measurements under consistent conditions for a larger library of substituents.
To overcome this, Sterimol parameters have been developed, which instead rely on geometric coordinates \cite{verloop1976development} and have proved useful in asymmetric catalysis \cite{harper2012multidimensional, sigman2016development}.
In contrast, Bragato \textit{et al.} \cite{bragato2020data} introduced a Hammett-inspired product model (HIP) able to fit parameters to diverse chemistries and properties without the need for external references or geometries. 
By establishing an internal reference, it also allows for the inclusion of diverse environments for each substituent in the fitting process, resulting in more balanced constants. 
Some of its successful applications in catalysis include the prediction of adsorption energies of small carbon molecules \cite{bragato2023occams}.

For further partitioning of substituent effects, we turn to combination rules.
The Hammett equation, detailed in the following section, was developed for singly-substituted compounds. 
However, we deal with complexes containing multiple ligands in organometallic catalysis, so access to the effect of each ligand and ligand combination is essential. 
Early studies on disubstituted and trisubstituted benzene derivatives using the Hammett equation indicated additive substituent effects under minimal steric inhibition of resonance \cite{Jaffé1953reexamination}.
Diverse combination rules have also been employed for estimating thermodynamic properties of mixtures \cite{al2004generating, DELHOMMELLE_2001_Inadequacy, fyta2012forcefield}. 
These include using the arithmetic mean for pure component properties to estimate collision diameters \cite{Lorentz_1881} and the geometric mean for potential well depths \cite{berthelot1898melange} in the Lennard-Jones potential.
The harmonic mean is used for second virial coefficients \cite{fender1962second}, and the sixth-power mean for rare gas systems \cite{waldman1993hagler}.\\
In this work, we introduce an approach that partitions the contributions of the metal and each ligand in organometallic catalysts using the SM reaction as a test case. 
This method facilitates computational ligand tuning through binding energy predictions and their implementation into volcano plots.
We assess and propose methods for retrieving and combining individual ligand effects that ensure statistically stable calculations.
The combining rule is integrated into a HIP model \cite{bragato2020data,bragato2023occams} (Fig. \ref{fig:workflow}b).
Utilizing a dataset of 25k oxidative addition relative binding energies, we also investigate the performance of this combination rule-enhanced HIP model (cHIP) as a baseline for $\Delta$-ML \cite{Ramakrishnan_deltaml}, which learns residuals and further mitigates excessive data needs. 
Subsequently, we show how the design flexibility afforded by cHIP can be used to expand a second, smaller catalyst dataset, DB2, into DB3 and conduct screening (Fig. \ref{fig:workflow}c). 


\section{Methods and computational details}
\subsection{HIP model}
In the context of homogeneous catalysis, assessing the binding free energies between catalysts and substrates at each step of the catalytic cycle aids in screening, as shown by Busch \textit{et al.}\cite{busch2015linear} and utilized later in this work.  
Herein, we multiply the Hammett equation \cite{hammett1935some, hammett1937effect} by $-RT$, where $R$ is the ideal gas constant and $T$ is the temperature, to approximate changes in binding  energy as a simple product,
\begin{equation}
    \Delta E_{lm} = -RT\mathrm{log}\left( \frac{K_{lm}}{K_{0m}} \right) \simeq \rho_m \sigma_{l}
    \label{eq:my_hammett}
\end{equation} 
where $\rho_m$ and $\sigma_{l}$ are constants for the metal $m$ and ligand group $l$, respectively. 
$K_{lm}$ and $K_{0m}$ are equilibrium constants, with the subscript $0$ designating a reference ligand group. 
In this work, a ligand group refers to all the ligands around one central metal in a complex, and furthermore, we also consider Hammett-based approximations of 
DFT-obtained relative binding energies, i.e.~neglecting all thermal contributions. \\
Although empirical $\sigma$ values for common substituents in the ionization of benzoic acid derivatives can be found in the literature, such parameters are absent for many ligands pertinent to homogeneous catalysis. 
Furthermore, caution is warranted when assuming the transferability of these values for chemistries of different natures \cite{pearson1952hammett}.
Note that in Eq. \ref{eq:my_hammett}, the necessity of establishing a ligand as a reference is due to the ligand space being larger compared to metals.\\
The HIP model, as introduced by Bragato \textit{et~al.} \cite{bragato2020data, bragato2023occams}, enables the investigation of similar reactions by extracting linear scaling factors between them and eliminating reliance on external references.
It comprises the following 3 steps:
\subsubsection{Model setup}
The model begins with an ansatz assuming that there exists an offset $\Delta E_{0m}$. 
Then each change in binding energy can be predicted with
\begin{multline}
    \Delta E_{lm} \simeq \rho_m \sigma_{l} + \Delta E_{0m}
    \label{eq:prediction}
\end{multline}
Furthermore, the binding energy changes can be stored in a matrix \textbf{M}, where $X$ is the number of metals and $Y$ is the number of ligand groups,
\begin{equation}
    \mathbf{M} = \begin{bmatrix}
        \Delta E_{11} & \Delta E_{12} & \hdots & \Delta E_{1m} & \hdots & \Delta E_{1X}\\
        \Delta E_{21} & \Delta E_{22} & \hdots & \Delta E_{2m} & \hdots & \Delta E_{2X}\\
        \vdots        & \vdots        & \ddots & \vdots        & \ddots & \vdots\\
        \Delta E_{l1} & \Delta E_{l2} & \hdots & \Delta E_{lm} & \hdots & \Delta E_{lX}\\
        \vdots        & \vdots        & \ddots & \vdots        & \ddots & \vdots\\
        \Delta E_{Y1} & \Delta E_{Y2} & \hdots & \Delta E_{Ym} & \hdots & \Delta E_{YX}
\end{bmatrix}
\label{M} 
\end{equation}

For each column, the internal reference $\Delta E_{0m}$ is defined as its median.

\subsubsection{Solving for $\rho$}
Hammett's equation (Eq. \ref{eq:my_hammett}) suggests a linear scaling relationship between energies of complexes with the same ligands but different metals.
An initial set of $\rho$s can then be the pairwise scaling factors $c_{mn}$ 
between any two columns $m$ and $n$ of \textbf{M}, thereby eliminating reference bias~\cite{bragato2020data}.
Consequently, $c_{mn} = \frac{\rho_{m}}{\rho_{n}} \simeq \frac{\Delta E_{lm}}{\Delta E_{ln}}$, or %
\begin{equation} 
     c_{mn}\rho_{n} - \rho_{m} =0
    \label{eq:rho_init}
\end{equation}

\begin{figure}[t]
\centering
    \includegraphics[scale=0.8]{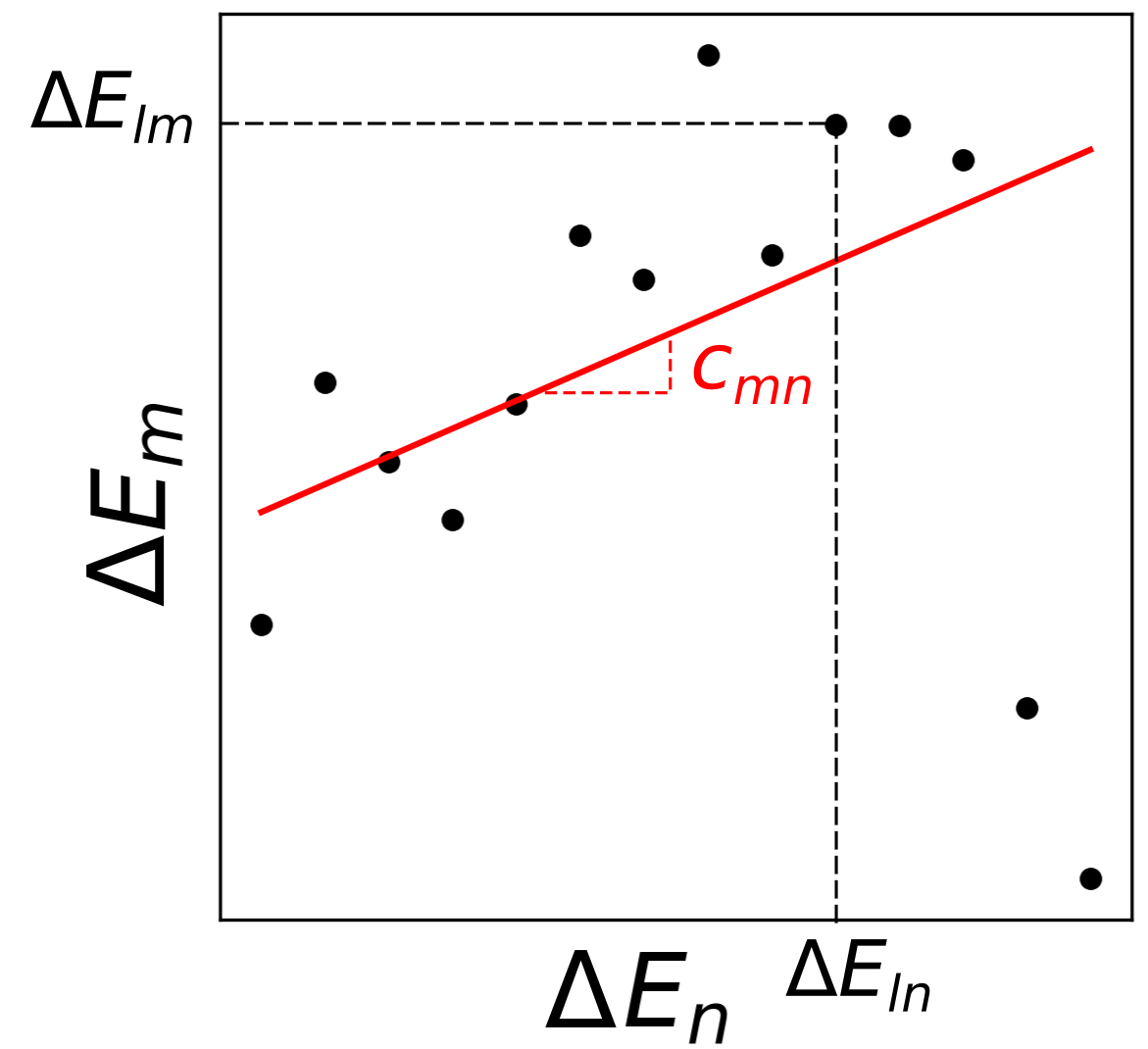}
    \caption{\justifying{Binding energies for complexes with metals $m$ and $n$. Each point represents a distinct ligand combination. The regression line, obtained through Theil-Sen regression~\cite{Theil_1950,sen1968estimates,bragato2020data}, has a slope representing the median of all pairwise slopes.}}
    \label{fig:theilsen}
\end{figure}

$c_{mn}$ is computed as the slope of the line of best fit between all $\Delta E_{lm}$ and $\Delta E_{ln}$ for $1 \leq l \leq Y$  (see Fig. \ref{fig:theilsen}). 
The procedure adopts Theil-Sen regression \cite{Theil_1950, sen1968estimates}, evaluating the median of $s=\binom{Y}{2}$ pairwise slopes $(\Delta E_{lm} - \Delta E_{km})/(\Delta E_{ln}-\Delta E_{kn})$ for $1 \leq l < k \leq Y$. 
Ordering the slopes in a list $S$, we get
\begin{equation}
  c_{mn}=\begin{cases}
    S_{\frac{s+1}{2}}, & \text{if $s$ is odd}.\\
    \frac{S_{\frac{s}{2}} + S_{\frac{s}{2}+1}}{2}, & \text{if $s$ is even}.
  \end{cases}
\end{equation}
This method offers the advantage of being more robust towards outliers, but it scales quadratically with the number of data points.

Accounting for all permutations of $m$ and $n$, $m \neq n$,  Eq. \ref{eq:rho_init} yields an overdetermined system of linear equations $\mathbf{C\boldsymbol{\rho}=0}$ which can be used to solve for $\rho$s.
A comprehensive description of this matrix is provided in the supplementary information section (SI).


\subsubsection{Solving for $\sigma$}
Subsequently, using the $\rho$s and matrix $\textbf{M}$ entries, we calculate $\sigma$ for each ligand group as
\begin{equation}
    \sigma_l = \frac{1}{X}\sum_{m=1}^{X} \frac{\Delta E_{lm}}{\rho_m}
\end{equation}
The $\rho$ of each metal is then refined via another Theil-Sen regression.
The slopes to evaluate in this case are $(\Delta E_{lm} - \Delta E_{km})/(\sigma_{l} - \sigma_{k})$ for $1 \leq l < k \leq Y$, and the new $\rho_m$ is the median slope.
After this refinement, a satisfying level of self-consistency is reached and further iterations are no longer necessary.
In this work, $\sigma$ is in energy units, and $\rho$ is considered a unitless pre-factor.

\subsection{Combination rule}
\begin{figure}
    \includegraphics[scale=0.45]{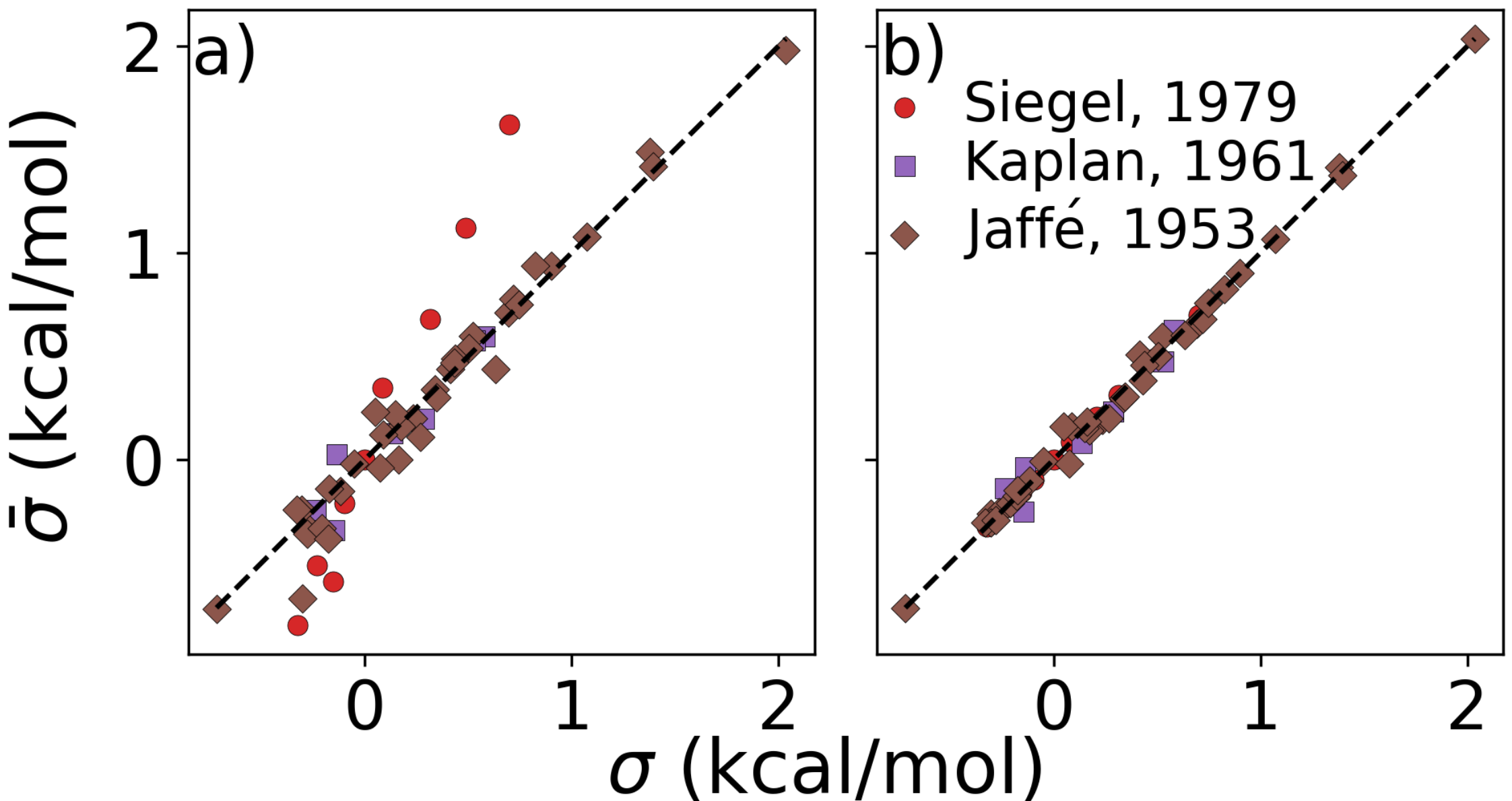}
    \caption{\justifying Test of combination rule for experimentally obtained substituent parameters describing reactions of various chemistries (see text) published decades ago by Siegel~\cite{siegel1979mechanism}, Kaplan~\cite{kaplan1961reactivity}, Jaff\'e\cite{Jaffé1953reexamination}. Substituent constants ($\bar{\sigma}$) were obtained by (a) summing published $\sigma$, and (b) after linear regression for each of the three published sets (this work).}
    \label{fig:least_squares}
\end{figure}
In cases where the $\sigma$s of individual substituents are known, one can assume that an additive combination rule can be used to retrieve the effects of any combination of them. This idea was generalized and confirmed already in 1953 by Jaff\'e who compiled overwhelming evidence in support of this effect \cite{Jaffé1953reexamination}. 
We note for this work that conversely individual ligand contributions can also be inferred as soon as sufficiently many $\sigma$s for combinations of substituents (groups) are known. We have exploited the latter idea to first quantify $\sigma$s for individual ligands via linear regression and to subsequently add them for estimating $\sigma$s of novel ligand combinations.

In order to reconfirm the validity of this approach we revisited the previously published data. In particular, the effect of a group of ligands $l$ is approximated as
\begin{equation}
    \sigma_{l} \simeq \sum_{i=1}^{N_i} \sigma_i
    \label{eq:jaffe}
\end{equation}
for a multisubstituted system. 
Here, $N_i$ is the number of ligands in the group $l$, and $\sigma_i$ is the effect of ligand $i$ in a monosubstituted system.
We estimate $\sigma_i$ with $\bar\sigma_i$ by solving the system of equations 
\begin{equation}
    \mathbf{D\boldsymbol{\bar\sigma=\sigma_l}}
    \label{eq:least_squares}
\end{equation}
where $D_{IJ}$ indicates the number of appearances of substituent $J$ in compound $I$, $\boldsymbol{\sigma_l}$ contains the sums of substituent effects for each compound.
$\boldsymbol{\bar\sigma}$ is the vector of single substituent effects that can be solved via the linear least squares method.

Results on display in Fig. \ref{fig:least_squares} confirm our expectation that
$\sigma$s of combinations are additive in $\sigma$s of single ligands. 
These experimental constants were obtained from studies on the hydrolysis of phosphonium salts \cite{siegel1979mechanism}, alcoholysis of isocyanates \cite{kaplan1961reactivity}, and various reactions involving benzoic acids \cite{Jaffé1953reexamination}.
The datasets encompassed di- and trisubstituted compounds featuring small substituents such as Cl, CH$_3$, OCH$_3$, NO$_2$, etc., each with established Hammett parameters. 
Furthermore, note that the sums of parameters obtained through linear regression (Fig.~\ref{fig:least_squares}b) 
are consistently closer to the experiments than the sums of Hammett parameters. 
This enhanced accuracy of the least squares method, presumably due to improved regularization and balancing, suggests its capability to robustly account for environment-specific synergistic effects, typically absent within the arbitrarily selected experiments resulting in the initial Hammett parameters.
We also report the performance of other combination rules in the SI, 
which revealed less accuracy than the additive rule. 

For estimating relative binding energies to catalyst complexes in the SM coupling reaction
we have first applied Eq. \ref{eq:least_squares} to  ligand pairs using the $\sigma$s obtained from HIP as $\boldsymbol{\sigma_l}$.
Then, cHIP predictions, $\Delta E^\mathrm{^c}$, can be obtained for any ligand pair $ij$ and catalyst metal $m$ via
\begin{widetext}
    \begin{eqnarray}
    \Delta E_{ijm} \simeq \Delta E_{ijm}^\mathrm{^c} = \rho_m \bar{\sigma}_{ij} + \Delta E_{\mathrm{0}m} = \rho_m (\sigma_i  + \sigma_j) + \Delta E_{\mathrm{0}m}
    \label{hammett_lorentz}
\end{eqnarray}
\end{widetext}

\subsection{$\mathrm{\Delta}$-ML}
To compare the above model to existing ML methods, we adopted an ML approach with Hammett to yield learning curves by fitting Hammett parameters to a growing training set and testing on fixed out-of-sample data.
To obtain predictions for all data points, the data was divided into 5 folds, and each was used as a test set once while training on the 4 others.
Due to varied training set sizes in the cHIP model, certain ligand combinations present in the test set might be absent from the training sets, especially with smaller training sets.
Hence, a categorical regression using one-hot encoding was employed to estimate $\sigma$ values for unknown ligands from known ones \cite{bragato2020data}. 
Ligand constants from this step were used only in cases where they could not be estimated during training.\\ 
The cHIP results served as a baseline for $\mathrm{\Delta}$-ML with KRR \cite{Ramakrishnan_deltaml}.
KRR, a supervised ML technique initially introduced in chemistry for learning molecular atomization energies \cite{hansen2013krr}, maps data from the input space to a feature space and calculates the dot product of the transformed vectors.
The mapped data in this case are the representations, which contain information derived from the molecular structure.
For a given set of $N$ training instances, the corrected predictions after $\mathrm{\Delta}$-ML are obtained using
\begin{equation}
    \Delta E\mathrm{^{\Delta}}= \Delta E\mathrm{^c}+\sum_{t=1}^{N} \alpha_{t}k(\mathbf{x}_t, \mathbf{x}_q)
\end{equation}
where the second term is the KRR-predicted residual.
$\mathbf{x}$ are the representations and $k$ is a similarity measure between the training compounds $t$ and the query compounds $q$, respectively. These similarity measures were obtained with a Laplacian kernel (Eq. \ref{laplacian}), or a Gaussian kernel (Eq. \ref{gaussian}) for computations using SLATM as a representation.
\begin{equation}
    k(\mathbf{x}_t, \mathbf{x}_q) = \mathrm{exp}\left( -\frac{||\mathbf{x}_t - \mathbf{x}_q||_1}{\sigma^{\prime}}\right) 
    \label{laplacian}
\end{equation}
\begin{equation}
    k(\mathbf{x}_t, \mathbf{x}_q) = \mathrm{exp}\left( -\frac{||\mathbf{x}_t - \mathbf{x}_q||_{2}^{2}}{{2\sigma^{\prime}}^2}\right) 
    \label{gaussian}
\end{equation}
Here, $\sigma^{\prime}$ is a hyperparameter optimized for every training set size using grid search, not to be confused with $\sigma$ in the Hammett equation.
$\boldsymbol{\alpha}$ is a vector of regression coefficients that is obtained using
\begin{equation}
    \boldsymbol{\alpha} = (\mathrm{\mathbf{K}} + \lambda \mathbf{\mathbf{I}})^{-1} \mathbf{\mathbf{y}}
\end{equation}
where $\lambda$ is a regularizer, optimized at every training set size, $\mathbf{I}$ the identity matrix, $\mathrm{\mathbf{y}}$ a vector containing the training properties, and $\mathrm{\mathbf{K}}$ a kernel matrix containing similarity measures between all compounds in the training set. 

\subsection{Datasets}
\renewcommand{\arraystretch}{1.3}
\begin{table*}
\caption{\justifying Overview of all datasets used and predicted in this work, including catalyst and reaction step details. All 108 molecular graphs of ligands are provided in the SI.}
    \begin{tabular}{ p{1cm} p{2.5cm} p{2cm}  p{1.5cm}  p{1.5cm}  p{3cm}  p{2cm} p{3cm}}
        \hline
        & \Centering\textbf{Purpose} & \Centering\textbf{Catalyst structure} & \Centering\textbf{Metals} & \Centering\textbf{$\#$ of Ligands} & \Centering\textbf{Reaction steps} & \Centering\textbf{$\#$ of data entries} & \Centering\textbf{Data origin}\\
        \hline
        \Centering\textbf{DB1} & \Centering Test combination rule & \Centering$L_{i}-M_m-L_{j}$ & \Centering Ni, Pd, Pt, Cu, Ag, Au & \Centering91, see SI & \Centering Oxidative addition & \Centering25116 & \Centering 7054 from DFT \cite{meyer2018} + 18062 from KRR using MBDF and Laplacian kernel (this work)\\ 
        \hline
        \Centering\textbf{DB2} & \Centering Catalyst discovery & \Centering$L_i-M_m-L_i$ & \Centering Ni, Pd, Pt, Cu, Ag, Au & \Centering16, see SI & \Centering Oxidative addition, transmetalation, reductive elimination & \Centering276 & \Centering276 from DFT (92 per reaction step) \cite{busch2017generalized}\\
        \hline
        \Centering\textbf{DB3} & \Centering Novel catalysts & $L_{i}-M_m-L_{j}$ & \Centering Ni, Pd, Pt, Cu, Ag, Au & 16, see SI & \Centering Oxidative addition, transmetalation, reductive elimination & 720 & cHIP (this work)\\
        \hline
    \end{tabular}
\label{tab:datasets}
\end{table*}
\begin{figure}[t]
    \includegraphics[scale=0.8]{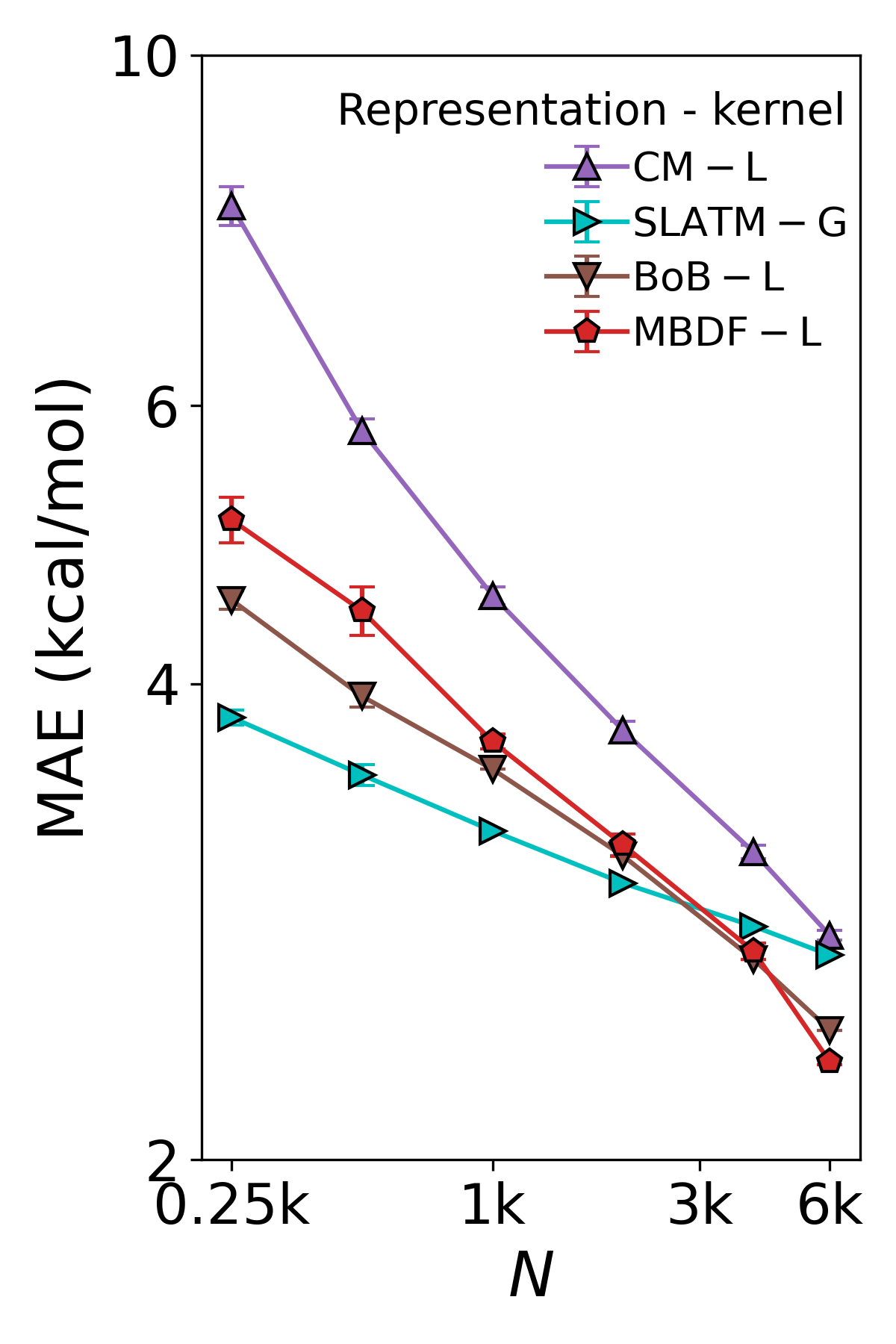}
    \caption{\justifying Test of ML model used to augment the DFT data in DB1 (Table ~\ref{tab:datasets}). Learning curves (test error of relative binding energies vs training set size) for oxidative addition for different ML methods. L: Laplacian kernel, G: Gaussian kernel. The augmentation was done using the MBDF-based model trained on all 7054 DFT training points published in Ref.~\cite{meyer2018}.
    }
    \label{fig:lc_ml}
\end{figure}

Two datasets, both containing relative binding energies relative to the formation of 1,3-butadiene depicted in Fig. \ref{fig:workflow}a, were used in this work, as summarized in Table \ref{tab:datasets}. \\
The first one, referred to as Database 1 (DB1) was obtained from Ref. \cite{meyer2018}. It contains a total of 91 ligands of types phosphines (P), N-heterocyclic carbenes (NHC), pyridines (Py), and other common ligands (Other).
Their chemical structures are provided in the SI.
Those ligands were combined with six transition metals (Ni, Pd, Pt, Cu, Ag, and Au) to form catalysts having the structure $L_{i}-M_m-L_{j}$, where $L_{i}$ and $L_{j}$ are ligands and $M_m$ is a metal, spanning a total set of 25116 compounds. 
It contained relative binding energies relative to the oxidative addition step in the SM C-C cross-coupling reaction depicted in Fig. \ref{fig:workflow}a for 4186 $L_{i}-L_{j}$ combinations, each coupled with all 6 metals.
The distributions of the energies by metal and by ligand type combinations, shown in Fig. S1 and S2 respectively, primarily show that the energies are more strongly correlated with metals than with ligands.\\ 
A subset of 7054 geometries in DB1 was optimized using the AiiDA automated platform \cite{aiida} at the B3LYP-D3/3-21G \cite{B3LYP2010, B3LYP2011} level of theory for the Ni, Pd, Cu, and Ag complexes, and B3LYP-D3/def2-SVP \cite{SVP} for the Pt and Au complexes in Gaussian09 \cite{g09}.
Subsequently, B3LYP-D3/def2-TZVP \cite{TZVP} single-point calculations were performed.
The remaining 18062 energies were predicted by us using a KRR model trained on the DFT energies using the Many-Body Distribution Functionals (MBDF) \cite{khan2023kernel} representation and a Laplacian kernel, providing full coverage of all ligand-metal combinations at a reduced computational cost.
Three other representations and kernel combinations were built for comparison, namely Coulomb Matrix \cite{rupp2012coulomb} and Bag of Bonds (BoB) \cite{BoB} with a Laplacian kernel, and Spectrum of London and Axilrod–Teller–Muto potential (SLATM) \cite{SLATM} with a Gaussian kernel, hence reproducing Meyer \textit{et al.}'s work \cite{meyer2018}.
As shown in Fig. \ref{fig:lc_ml}, the comparison leads to the conclusion that MBDF, with the Laplacian kernel, yields the best result.\\
The second and smaller dataset (Database 2 or DB2) was obtained from Ref. \cite{busch2017generalized} and provides relative binding free energies for all 3 intermediate steps depicted in Fig. \ref{fig:workflow}a. 
The values include unscaled enthalpies and vibrational entropy contributions.
It contains symmetrical complexes, composed of the same 6 metals as in the first dataset and 16 ligands.
10 out of those 16 ligands were also present in the first one, and the energies of 4 Cu-based complexes were missing, yielding only 92 complexes per reaction step. 
The geometries of these complexes were optimized using M06/def2-SVP \cite{zhao2008density, zhao2008M06, SVP} in Gaussian09 \cite{g09} while accounting for solvation in tetrahydrofuran using the implicit SMD model \cite{marenich2009SMD}.

\section{Results and discussion}
\subsection{cHIP on oxidative addition (DB1)}
\begin{figure}[h]
    \includegraphics[scale=0.35]{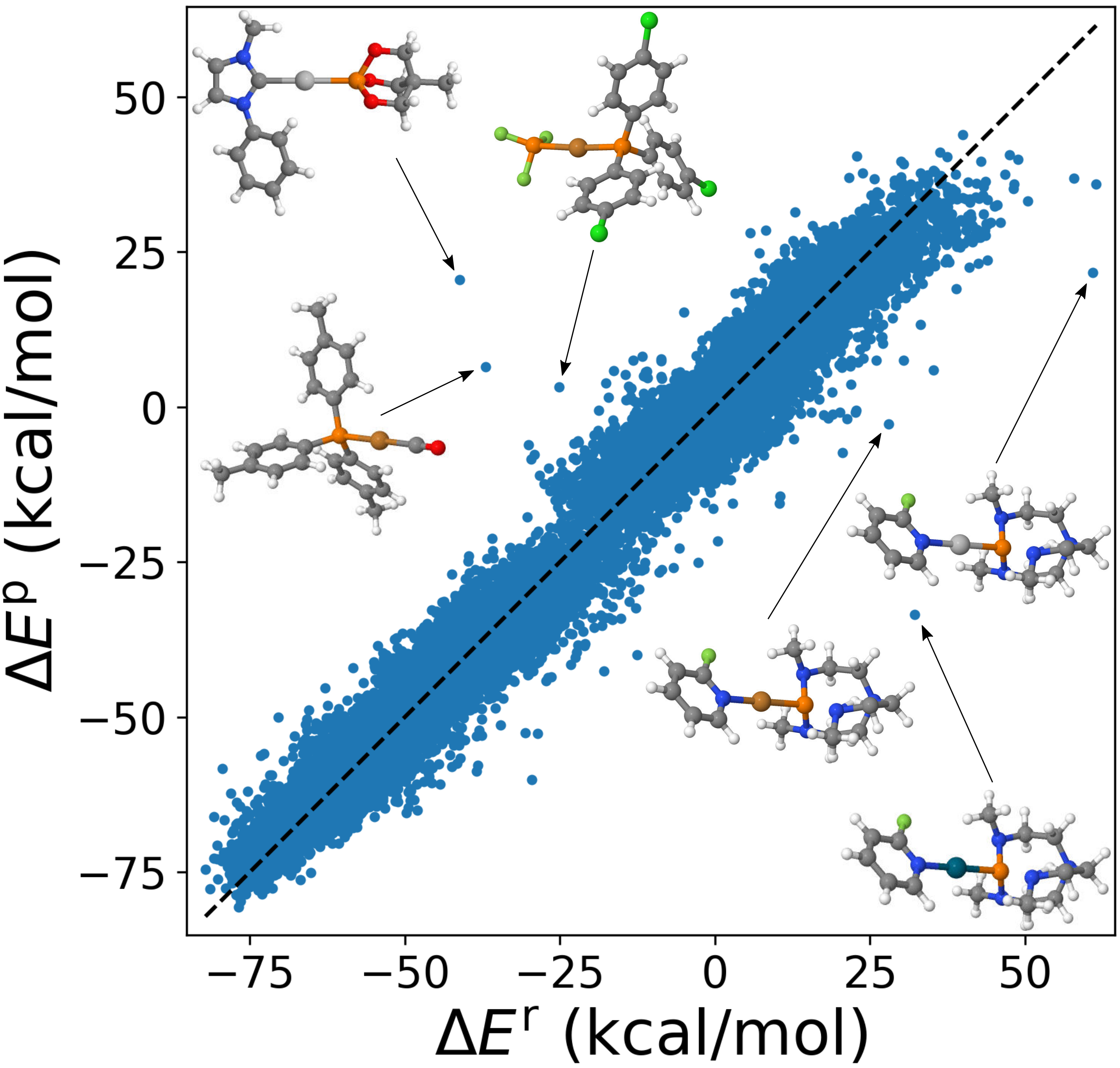}
    \caption{\justifying
   cHIP-predicted changes in binding energies against reference energies for DB1 (25k compounds) where MAE = 3.4 kcal/mol. Insets display complexes that deviate most (MAE $>$ 30 kcal/mol). Atom colors are gray, blue, white, green, red,  orange, silver, dark cyan, and dark orange for C, N, H, Cl, O, P, Ag, Pd, and Cu, respectively.}
    \label{fig:parity_plot}
\end{figure}

Application of the combination rule to 91 substituent specific $\sigma$s within cHIP (Eq. \ref{hammett_lorentz}) resulted in a mean absolute error (MAE) of $\sim$3.4 kcal/mol for DB1. 
We note that the naive HIP model (Eq. \ref{eq:prediction}), which accounts for changes in binding using 91$^2$/2 = 4186 global $\sigma$s, 
reached a MAE of $\sim$2.5 kcal/mol.
This increase of MAE from HIP to cHIP is expected due to the decrease in dimensionality from the additional layer of approximation introduced by estimating the combined ligand contributions as the sum of individual ones.
Corresponding scatter plots of 25116 model predictions versus reference data numbers (used for fitting) are shown in Figs. \ref{fig:parity_plot} and S3 for cHIP and HIP, respectively. 
Such predictive power, only on the order of a few percentage points of the range of the property (see Fig.~S1) is promising. We note that it is on par with popular density functional approximations, 
as well as with previous HIP results obtained for estimating activation energies in S$_N$2 reactions~\cite{bragato2020data}. 

Note how in Fig.~\ref{fig:parity_plot}, 
no skewing is observed (see also the error distribution curve per metal in Fig. S5).
As displayed as insets, outliers correspond to varying metals. 
However, they all have in common that there is one P ligand. 
Furthermore, all underestimated outliers correspond to catalysts that share the same ligands, proazaphosphatrane, and 2-fluoropyridine.

Most outliers of the cHIP model (Fig. \ref{fig:parity_plot}) 
were also outliers already for the HIP model (Fig.~S3). 
This suggests that, beyond the decrease in dimensionality caused by the combination rule, these shortcomings are likely to be caused by the Hammett equation's inadequacy in describing those specific ligands.
This inadequacy can only be partly explained by steric hindrance since the dataset also comprised several bulky NHC ligands which were not outliers.
We also note that since many systems used for fitting are sterically hindered, some steric effects could be included in the cHIP parameters.

Performing the regression on subsets according to the categories of each of its ligands further improved the prediction accuracy. 
In Fig. \ref{fig:learning_curves_all}, all the cHIP models fitted on the subsets had lower errors than the full dataset model.
These observations agree with the statement that a Hammett correlation occurs between closely related species \cite{Chattopadhyay_helper_1976}. 
They are also in line with the fact that regression is more difficult the higher the dimensionality~\cite{lemm2023improved}.
As such, cHIP promises to be a useful model for combinatorially scaling spaces (the combinations of ligands in this case) with low-dimensional chemistry-specific properties, and where limited training instances are available.

\begin{figure}[t]
    \centering
    \includegraphics[scale=0.48]{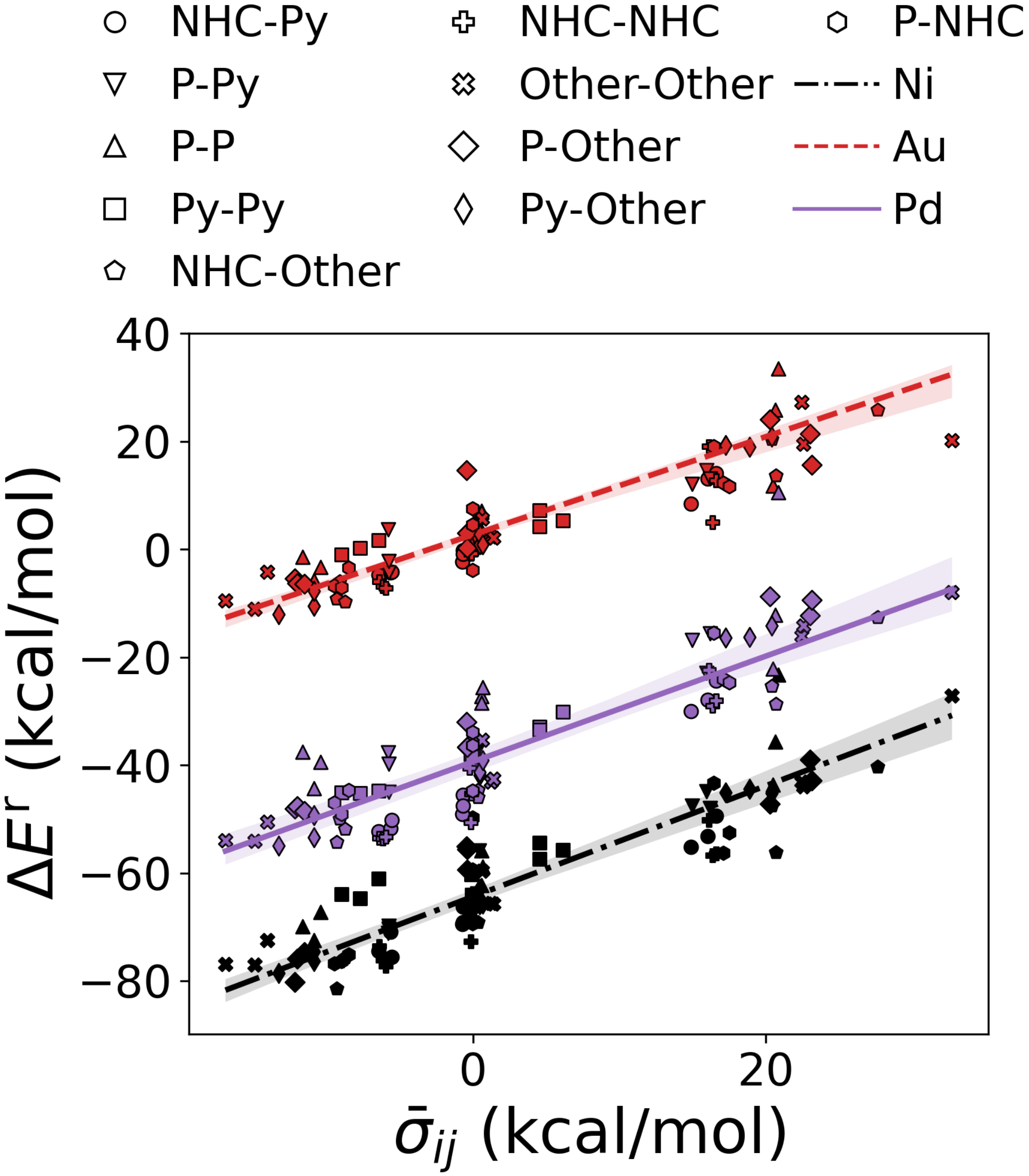}
    \caption{\justifying 
    Demonstration of cHIP.  Oxidative addition relative binding energies (DB1, see Table ~\ref{tab:datasets}) are shown as a function of averaged $\sigma$ obtained from cHIP. For the sake of simplicity, only 5 examples from each ligand type combination (given as legends) are plotted for each of the three metals. Lines correspond to cHIP predictions.} 
    \label{fig:hammettplot}
\end{figure}

The individual $\sigma$s were retrieved by applying the combination rule on the HIP predictions to produce the Hammett plot in Fig. \ref{fig:hammettplot}.
Moreover, the distinct clustering of complexes by their central metal and the linear trends in Fig. \ref{fig:hammettplot} validate the adequacy of the partitioning of $\rho$ and $\sigma$.

\begin{figure}[t]
    \centering
    \includegraphics[scale=0.57]{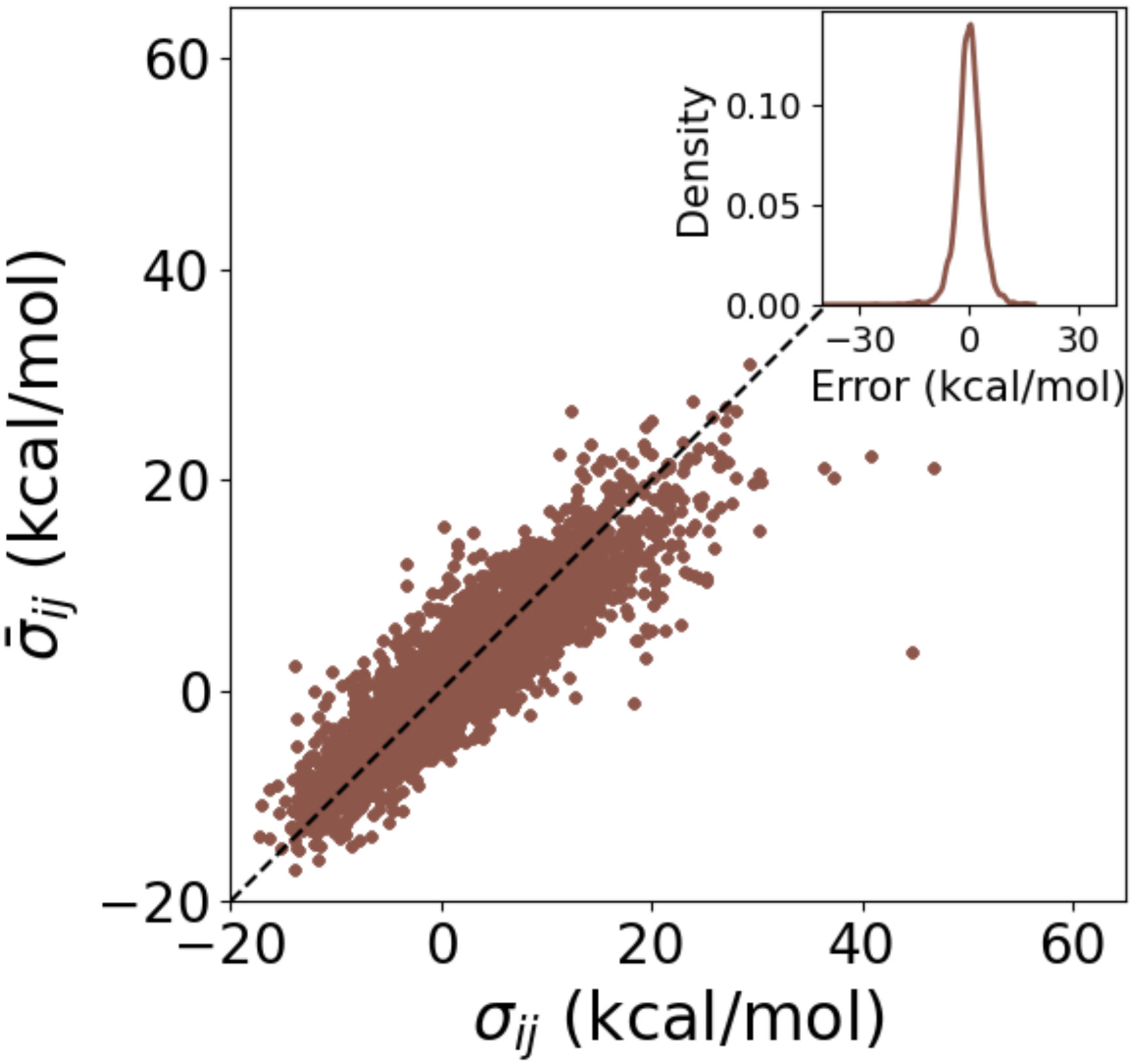}
    \caption{\justifying 
    Test of combination rule for individually
obtained substituent parameters using DB1 (Table \ref{tab:datasets}). 
Averaged $\bar{\sigma}$ (cHIP on DB1) vs reference $\sigma$ (HIP on DB1) for combinations of ligands $i$ and $j$ . 
The inset shows the corresponding error distribution. 
    }
    \label{fig:combining_rule}
\end{figure}

A closer look at the calculation of the individual $\sigma$s for DB1 in Fig. \ref{fig:combining_rule} revealed a MAE of 2.5 kcal/mol between the $\sigma$s from HIP and cHIP.
While an overestimation is usually observed with the additive rule when summing existing Hammett parameters \cite{Jaffé1953reexamination,Bhasha2009quantification}, we have circumvented this by fitting those parameters with linear regression.
The corresponding error distribution of cHIP $\sigma$s (inset of Fig~\ref{fig:combining_rule}) illustrates the welcome absence of any bias. 
\\

\subsection{$\Delta$-ML}
\begin{figure}[t]
    \centering
    \includegraphics[scale=0.7]{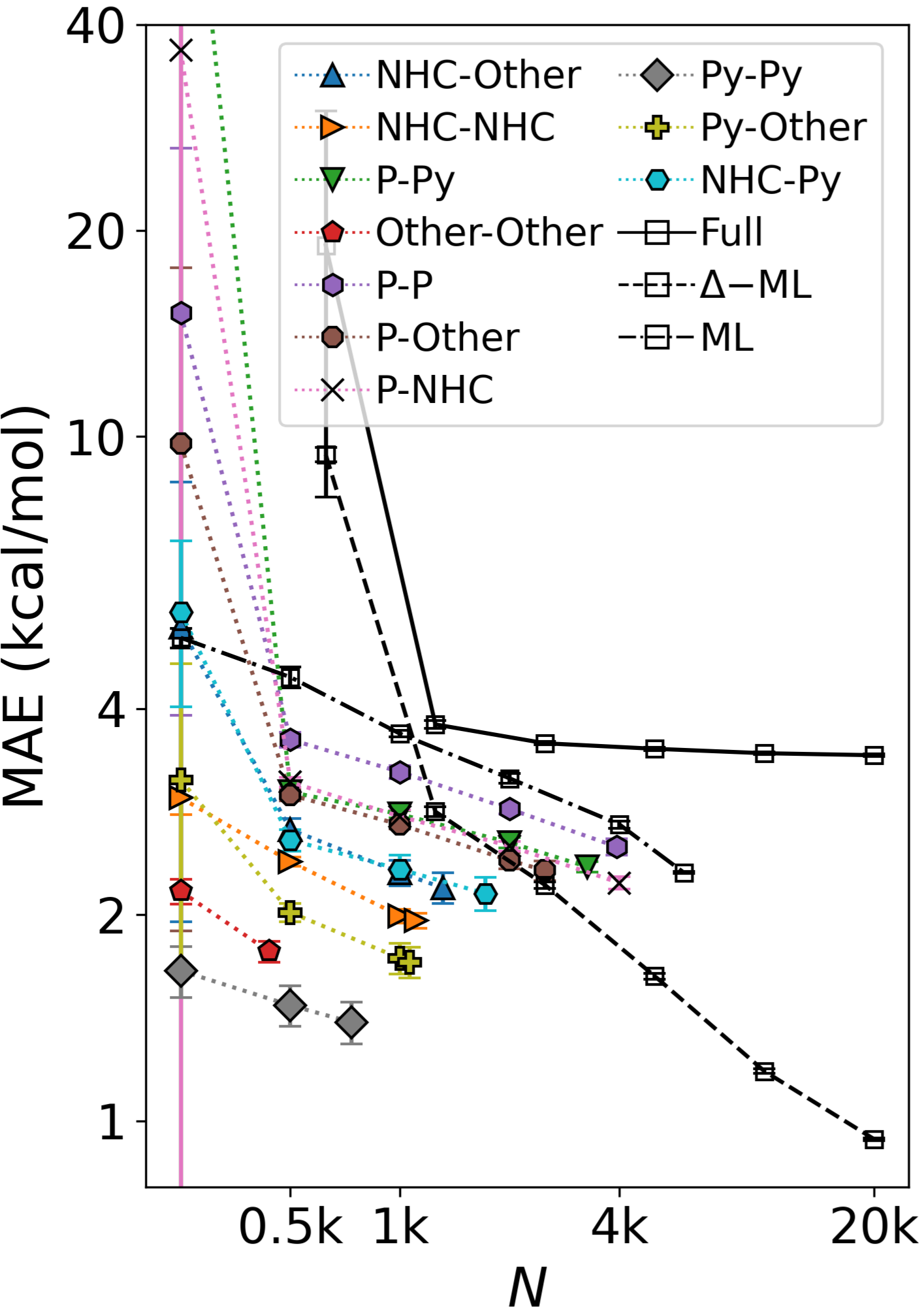}
    \caption{\justifying Mean absolute error of predicting relative binding energies in the oxidative addition step as a function of training set size. Errors of cHIP models on subsets (colored) and full DB1 (Table ~\ref{tab:datasets}) plateau rapidly. 
    ML corresponds to the KRR/MBDF line shown in Fig.~\ref{fig:lc_ml}. 
      $\Delta$-ML model trained on DB1 results with cHIP baseline results in lower offset and enables convergence to chemical accuracy.
      Error bars indicate standard deviations.}
    \label{fig:learning_curves_all}
\end{figure}

When regressing cHIP onto the entire DB1 (DFT + ML augmented instances), 
the prediction error ceases to improve, leveling off at $\sim$4 kcal/mol (see Fig. \ref{fig:learning_curves_all}). 
Note how the offset of the plateau is located at $\sim$1k training instances which roughly corresponds to the total number of parameters to fit in Eq. \ref{hammett_lorentz}.
While several physics-based representations, such as CM \cite{rupp2012coulomb}, global MBDF \cite{khan2023kernel}, BoB \cite{BoB}, or SLATM \cite{SLATM}, enable KRR models to systematically improve with training set size (see Fig. \ref{fig:lc_ml}), 
DB1 does not contain sufficient DFT instances to allow for convergence lower than 2 kcal/mol. 
Understanding catalytic yield relies on thermodynamic quantities for homogeneous catalysis \cite{busch2015linear}, where equilibrium constants exhibit exponential scaling with free energy differences, underscoring the significance of predicting such changes with chemical accuracy (1 kcal/mol).
Encouragingly, when combining cHIP as a baseline with a machine-learned 
correction, $\Delta$-ML, the resulting learning curve on DB1 continues to improve with training set size, reaching a MAE corresponding to chemical accuracy for $\sim$20k training instances. 
The systematically decreasing standard deviations of the prediction errors in the learning curves are equally promising, and can be attributed to an increasingly slimmer error distribution, similar to previously noted trends for ML models of formation energies of crystals~\cite{Elpasolite_2016}. 

\subsection{Catalyst discovery (DB2)}
\begin{figure}[t]
    \centering
    \includegraphics[scale=0.4]{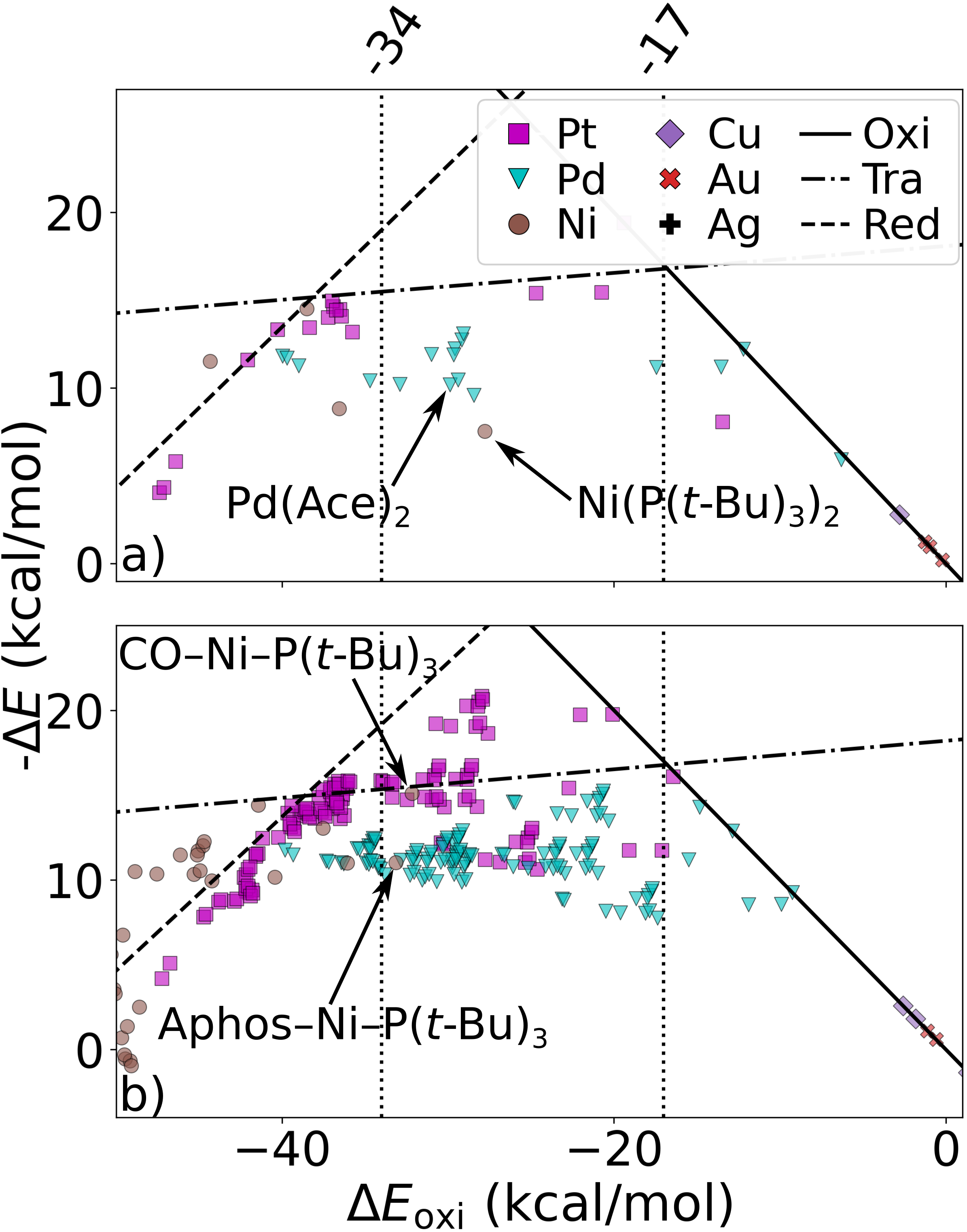}
    \caption{\justifying Volcano plot: Negative relative binding free energies of each step are plotted as a function of that of the oxidative step (oxi).  Ideal catalysts lie at the top of the volcano and vertical lines represent the ideal range as identified by Busch \textit{et al.} \cite{busch2017generalized}. a) catalysts sourced from DB2 (Table~\ref{tab:datasets}). b) novel catalysts predicted using cHIP, the 198 most interesting catalysts lie in between the two intersections of the transmetalation with either oxidative addition or reductive elimination. Two inexpensive (Ni-based) catalysts in the optimal regime are indicated by arrows. } 
    \label{fig:volcanoplot}
\end{figure}
We have studied the utility of cHIP in catalyst discovery through volcano plots.
For homogeneous catalysis, such plots are built by establishing linear scaling relationships between all the intermediate steps against a reference step \cite{busch2015linear} using a descriptor such as the catalyst-substrate binding free energy.
This allows for identifying an ideal range of energies in which only certain catalysts fall, aligning with Sabatier's principle \cite{sabatier1913catalyse}.
This strategy provides a way of screening potential catalysts without kinetic data.
DB2 was used in this phase as it contains relative binding free energies for all 3 steps.

The fitting of HIP to the entire dataset, illustrated in Fig. \ref{fig:volcanoplot}a, resulted in MAEs consistently below 3.5 kcal/mol for all steps.
Leveraging the effects of 16 single ligands, cHIP successfully predicted the ligand effects of 120 new ligand combinations.
Pairing these with each of the 6 metals, we predicted relative binding free energy changes for an additional 720 catalysts, reported in DB3.

As depicted in Fig. \ref{fig:volcanoplot}b, 198 of the new catalysts lie at the top of the predicted volcano plot.
Among these, 145 displayed oxidative addition relative binding free energies ranging from -34.0 to 17.0 kcal/mol, previously identified as an optimal range by Busch \textit{et al.} \cite{busch2017generalized}. 
This combinatorial approach revealed several Ni-based catalysts approaching the top of the volcano after ligand tuning, despite the initially strong-binding nature of Ni. 
The discovery of these catalysts, derived from a metal that is more cost-effective and earth-abundant than the prominent Pd, is particularly attractive and has been actively explored in the field over the past decade \cite{han2013Pd_to_Ni, baviskar2023recent}.
When considering the cost of the ligands, 
the most cost-effective catalyst identified by cHIP is Aphos-Ni-P(\textit{t}-Bu)$_3$, representing about 67\% of the cost of the least expensive catalyst found in DB2, Pd(Ace)$_2$, based on ligand and metal prices provided by Sigma-Aldrich \cite{MilliporeSigma}. 
Similar phosphine ligands for Ni catalysts have been reported in literature such as ProPhos \cite{yang2024prophos} or P(Cy)$_3$ \cite{chen2011nickel}.

Notably, the model was able to capture a reasonable trend across both datasets, despite DB1 lacking solvent effects and DB2 incorporating implicit solvation. 
This suggests that the cHIP model could potentially make equally reliable predictions when trained on datasets that include explicit solvation effects.

\section{Conclusion}
Our study demonstrates the efficacy of employing an additive combination rule with Hammett's equation for computational catalyst discovery. We have exemplified its applicability to the prediction and analysis of organometallic complexes relevant to the catalysis of the Suzuki-Miyaura (SM) cross-coupling reaction. 
We exploit fitted Hammett parameters obtained from linear least squares regression across all available ligand combinations, surpassing the prediction performance of published Hammett parameters.
Due to its simplicity, this approach offers significant computational advantages, obviating the need for geometric coordinates or extensive computing resources, therefore serving as a quick yet useful screening tool.
It also proves valuable when dealing with smaller chemical compound subspaces. 
Furthermore, our findings illustrate its utility as a baseline for $\Delta$-ML, reaching predictive power with chemical accuracy. 
The ability to separate ligand pair effects into single ligand effects facilitates the exploration of larger catalyst spaces and ligand tuning. 
This was demonstrated for a dataset comprising symmetrical catalysts created through the combination of 16 ligands and 6 metals.
Our model identified 145 new catalyst candidates for the SM reaction, including some based on Ni implying potentially substantial cost savings. 
For example, we identified the phosphine ligand-based Ni catalyst, Aphos-Ni-P($t$-Bu)$_3$, which seems to fall in line with other reported phosphine ligands. 
Future research directions will include further investigation of the effect of crowding and the specific environments on the ligand constants, aiming to elucidate the outliers in the model (phosphorus-containing ligands) and extend this approach to other catalysts and complexes with more than two ligands.

\section*{Concflicts of interest}
There are no conflicts to declare.

\begin{acknowledgments}
We acknowledge the support of the Natural Sciences and Engineering Research Council of Canada (NSERC), RGPIN-2023-04853.
O.A.v.L. has received funding from the European Research Council (ERC) under the European Union’s Horizon 2020 research and innovation programme (grant agreement No. 772834).
This research was undertaken thanks in part to funding provided to the University of Toronto's Acceleration Consortium from the Canada First Research Excellence Fund,
grant number: CFREF-2022-00042.
O.A.v.L. has received support as the Ed Clark Chair of Advanced Materials and as a Canada CIFAR AI Chair.  \end{acknowledgments}

\bibliographystyle{rsc}
\bibliography{main1}

\end{document}


\title{Supplementary information: Combining Hammett $\sigma$ constants for $\Delta$-machine learning and catalyst discovery}
\author{V. Diana Rakotonirina, Marco Bragato, Stefan Heinen, O. Anatole von Lilienfeld}
\date{}
\maketitle

\section{Construction of the system of linear equations}

\renewcommand{\theequation}{S\arabic{equation}}
To solve for the metal constants $\mathbf{\boldsymbol{\rho}}$, the Hammett-inspired product model (HIP) follows the blueprint laid out in Eq. 4 to build a system of linear equations $\mathbf{C\boldsymbol{\rho}=0}$. 
Accounting for all permutations of $m$ and $n$, $m \neq n$,  this gives us

\begin{equation}\label{eq:rho_matrix}
\resizebox{0.75\hsize}{!}{%
    $\begin{blockarray}{cccccccc}
    \begin{block}{c[ccccccc]}
    m=1, n=2 & -1 & c_{12}&0&\hdots&0&0&0\bigstrut[t]\\
    m=1, n=3 & -1 & 0 & c_{13} &\hdots&0&0&0\\
    \vdots&\vdots & \vdots & \vdots & \ddots & \vdots & \vdots & \vdots\\
    m=1, n=X-1 & -1 & 0 & 0      &\hdots  & 0&c_{1X-1}&0\\
    m=1, n=X & -1 & 0 & 0      &\hdots  & 0&0&c_{1X}\\
    m=2, n=1 & c_{21} & -1     & 0      & \hdots & 0      & 0      & 0\\
    m=2, n=3 & 0      & -1     & c_{23} & \hdots & 0      & 0      & 0\\
    \vdots&\vdots & \vdots & \vdots & \ddots & \vdots & \vdots & \vdots\\
    m=2, n=X-1 & 0      & -1     & 0      & \hdots & 0      & c_{2X-1} & 0\\
    m=2, n=X & 0      & -1     & 0      & \hdots & 0      & 0      & c_{2X}\\
    m=3, n=1 & c_{31} & 0      & -1     & \hdots & 0      & 0      & 0 \\
    m=3, n=2 & 0      & c_{32} & -1     & \hdots & 0      & 0      & 0 \\
    \vdots&\vdots & \vdots & \vdots & \ddots & \vdots & \vdots & \vdots\\
    m=X, n=X-2 & 0      & 0      & 0      & \hdots & c_{XX-2} & 0    & -1\\
    m=X, n=X-1 & 0      & 0      & 0      & \hdots & 0      &c_{XX-1} & -1  \bigstrut[b]\\
    \end{block}
    \end{blockarray}
\begin{bmatrix}
    \rho_1\\
    \rho_2\\
    \rho_3\\
    \vdots\\
    \rho_{X-2}\\
    \rho_{X-1}\\
    \rho_X
\end{bmatrix}
= \begin{bmatrix}
    0\\
    0\\
    \vdots\\
    0\\
    0\\
    0\\
    0\\
    \vdots \\
    0 \\
    0\\
    0\\
    0\\
    \vdots\\
    0\\
    0\\
\end{bmatrix}$%
}
\end{equation}

Constraining the first component of the vector of $\rho$s to be 1 results in a vector \textbf{r} of length $(X-1)$ containing the remaining $\rho$ values.
HIP thus avoids trivial solutions, transforming the homogeneous system in Eq. \ref{eq:rho_matrix} into $\mathbf{C'r=c}$ (Eq. \ref{eq:rho_matrix2}), where \textbf{c} is the first column of the matrix in Eq. \ref{eq:rho_matrix} and \textbf{C'} is the remaining part.
The equation to be solved for $\rho$s then becomes
\begin{equation}\label{eq:rho_matrix2}
\resizebox{0.75\hsize}{!}{%
    $\begin{blockarray}{cccccccc}
    \begin{block}{c[ccccccc]}
    m=1, n=2 & c_{12}&0&\hdots&0&0&0\bigstrut[t]\\
    m=1, n=3 & 0 & c_{13} &\hdots&0&0&0\\
    \vdots & \vdots & \vdots & \ddots & \vdots & \vdots & \vdots\\
    m=1, n=X-1 & 0 & 0      &\hdots  & 0&c_{1X-1}&0\\
    m=1, n=X& 0 & 0      &\hdots  & 0&0&c_{1X}\\
    m=2, n=1 & -1     & 0      & \hdots & 0      & 0      & 0\\
    m=2, n=3 & -1     & c_{23} & \hdots & 0      & 0      & 0\\
    \vdots& \vdots & \vdots & \ddots & \vdots & \vdots & \vdots\\
    m=2, n=X-1 & -1     & 0      & \hdots & 0      & c_{2X-1} & 0\\
    m=2, n=X  & -1     & 0      & \hdots & 0      & 0      & c_{2X}\\
    m=3, n=1& 0      & -1     & \hdots & 0      & 0      & 0 \\
    m=3, n=2 & c_{32} & -1     & \hdots & 0      & 0      & 0 \\
    \vdots& \vdots & \vdots & \ddots & \vdots & \vdots & \vdots\\
    m=X, n=X-2  & 0      & 0      & \hdots & c_{XX-2} & 0    & -1\\
    m=X, n=X-1 & 0      & 0      & \hdots & 0      &c_{XX-1} & -1  \bigstrut[b]\\
    \end{block}
    \end{blockarray}
\begin{bmatrix}
    \rho_2\\
    \rho_3\\
    \vdots\\
    \rho_{X-2}\\
    \rho_{X-1}\\
    \rho_X
\end{bmatrix}
= \begin{bmatrix}
    1\\
    1\\
    \vdots\\
    1\\
    1\\
    -c_{21}\\
    0\\
    \vdots \\
    0 \\
    0\\
    -c_{31}\\
    0\\
    \vdots\\
    0\\
    0\\
\end{bmatrix}$%
}
\end{equation}
\section{Other combination rules}
The other combination rules that were compared to the additive one are geometric averaging (Eq. \ref{eq:geometric}) [1], a sixth-power average proposed by Waldman and Hagler (Eq. \ref{eq:waldman}) [2], and a harmonic mean proposed by Fender and Halsey (Eq. \ref{eq:fender}) [3].
\begin{equation}
    \sigma_{l} = \epsilon\sqrt{\prod_{i=1}^{N_i}\sigma{i}}
    \label{eq:geometric}
\end{equation}
\begin{equation}
    \sigma_{l} = \epsilon \left( \frac{\sum_{i=1}^{N_i}\sigma_{i}^6}{N_i} \right)^{1/6}
    \label{eq:waldman}
\end{equation}
\begin{equation}
    \sigma_{l} = \epsilon\frac{N_i\prod_{i=1}^{N_i}\sigma{i}}{\sum_{i=1}^{N_i}\sigma_{i}}
    \label{eq:fender}
\end{equation}
Given that the relative binding energies can be negative, another parameter $\epsilon$ was added, which had the value 1 if $\sigma_{l}$ was positive, and -1 otherwise. 
Results of their test are shown in the next section.

\section{Supplementary figures}
\begin{figure}[H]
    \centering
    \includegraphics[width=0.48\textwidth]{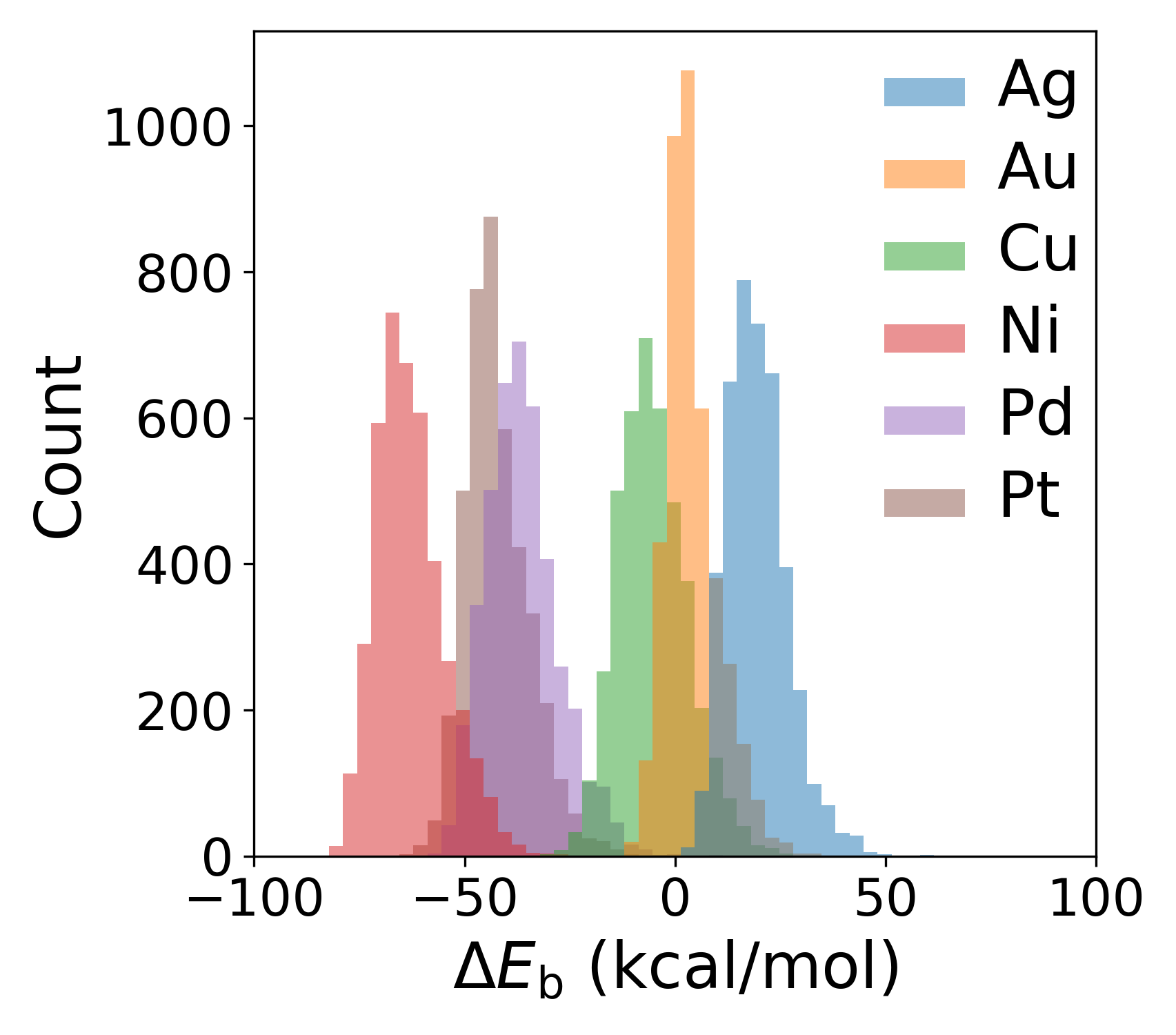}
    \caption{Distribution of relative binding energies by metal for all 25k catalysts in DB1.}
    \label{fig:metal_dist}
\end{figure}

\begin{figure}[H]
    \centering
    \includegraphics[width=0.48\textwidth]{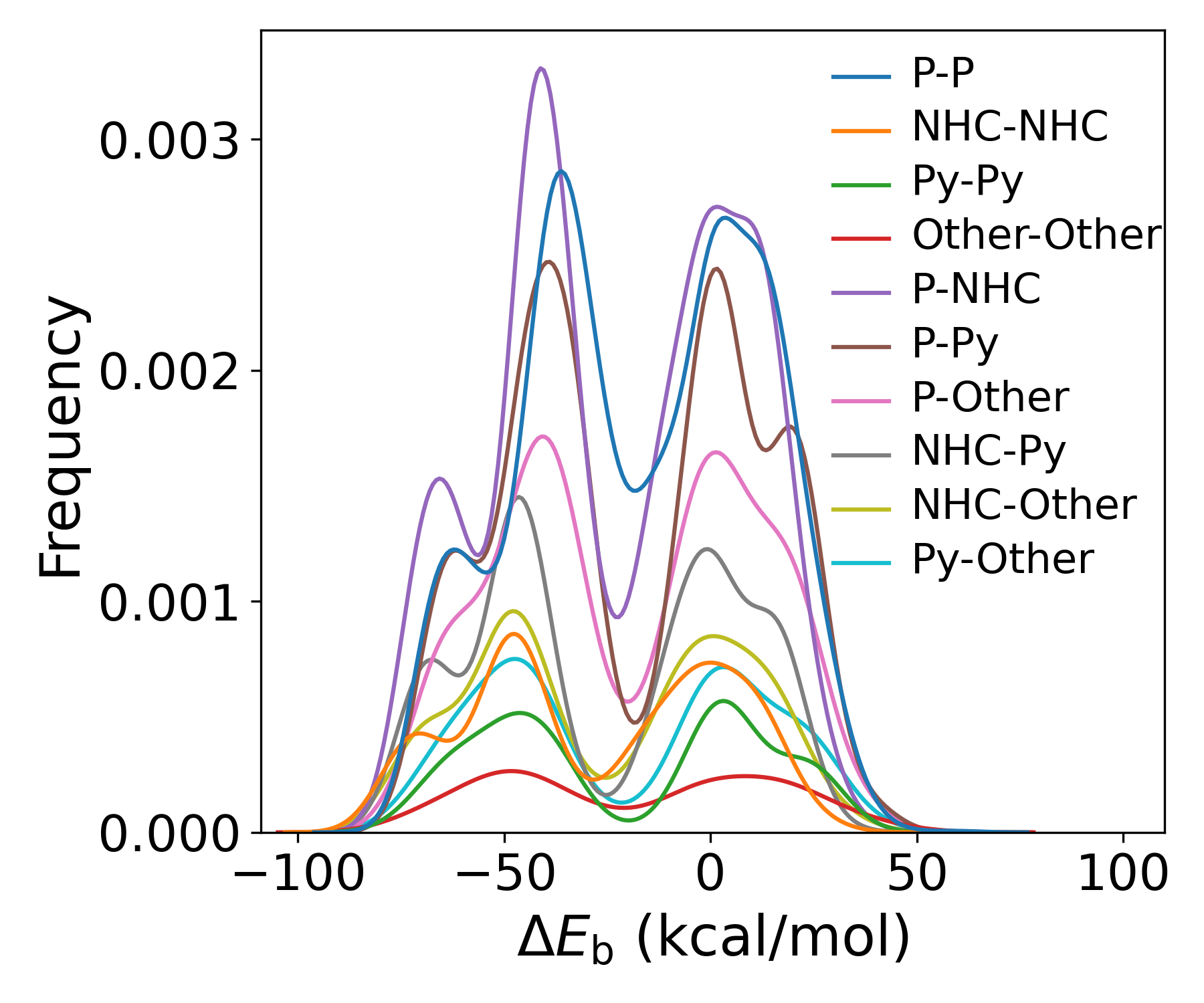}
    \caption{Kernel density estimation showing distribution of relative binding energies by ligand pair ($L_i-L_j$). Note that $L_i-L_j$ is considered the same as $L_j-L_i$.}
    \label{fig:ligand_dist}
\end{figure}




\begin{figure}[H]
    \centering
    \includegraphics[scale=0.25]{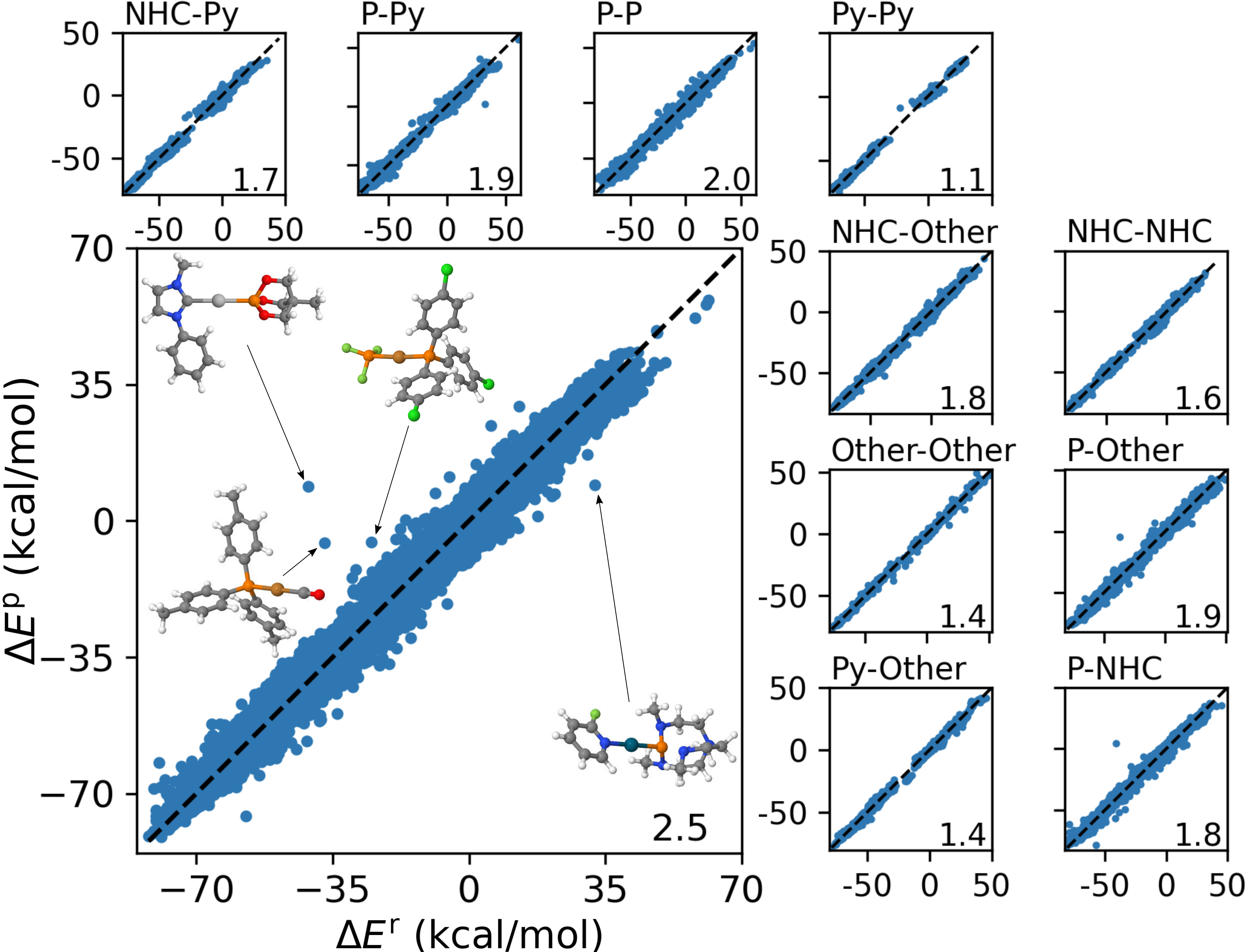}
    \caption{\justifying HIP-predicted relative binding energies against reference energies for full DB1 (bottom left) and subsets based on ligand type combinations (surroundings). Insets display complexes that deviate most (MAE $>$ 18 kcal/mol). Atom colors are gray, blue, white, green, red,  orange, silver, dark cyan, and dark orange for C, N, H, Cl, O, P, Ag, Pd, and Cu, respectively.}
    \label{fig:parity_plot}
\end{figure}



\begin{figure}[H]
    \centering
    \includegraphics[scale=0.6]{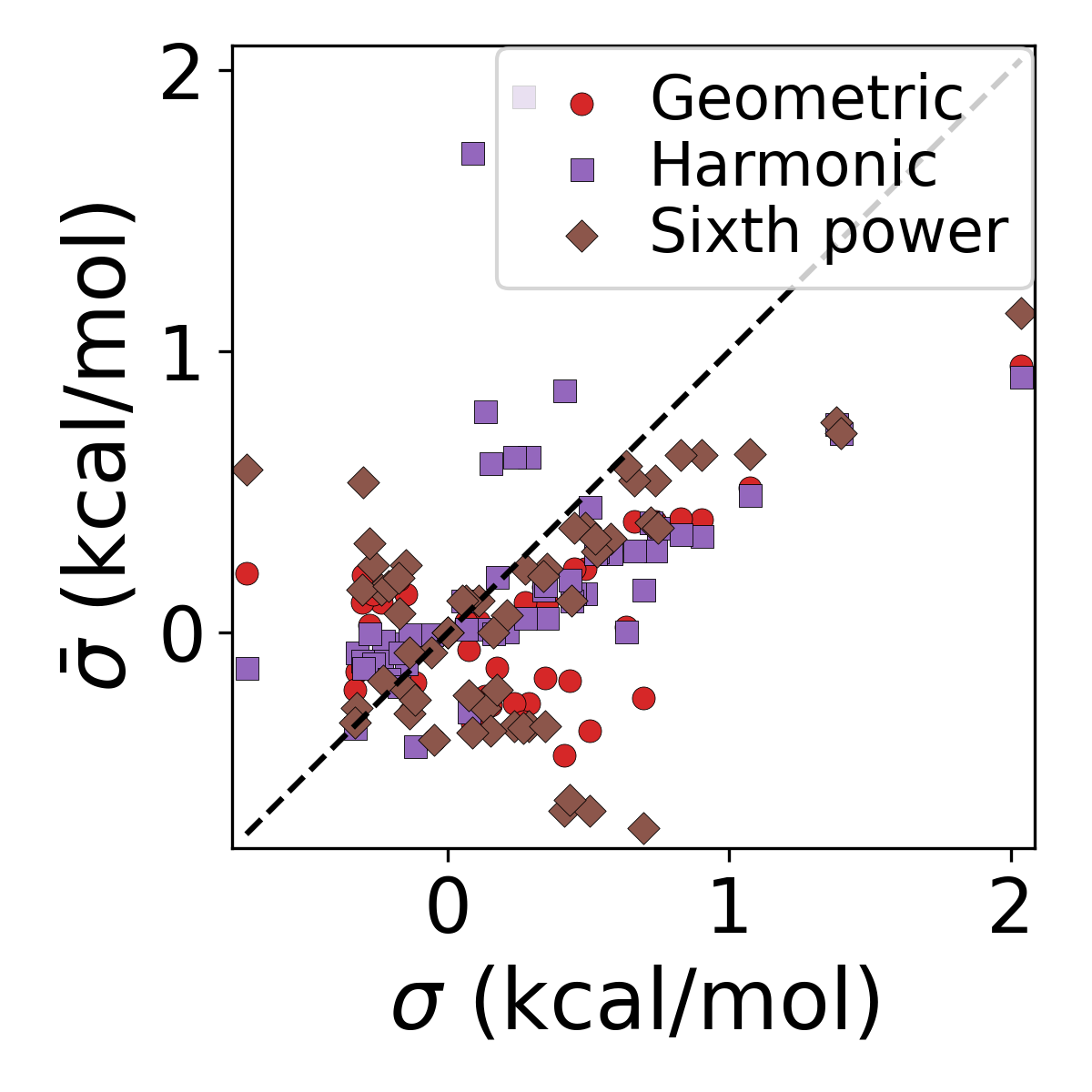}
    \caption{\justifying Test of different combination rules for experimentally obtained substituent parameters describing reactions of various chemistries (see main text). Substituent constants were obtained by combining published Hammett constants according to the rules in section S2.}
    \label{fig:sigma_all}
\end{figure}


\begin{figure}[H]
    \centering
    \includegraphics[width=0.48\textwidth]{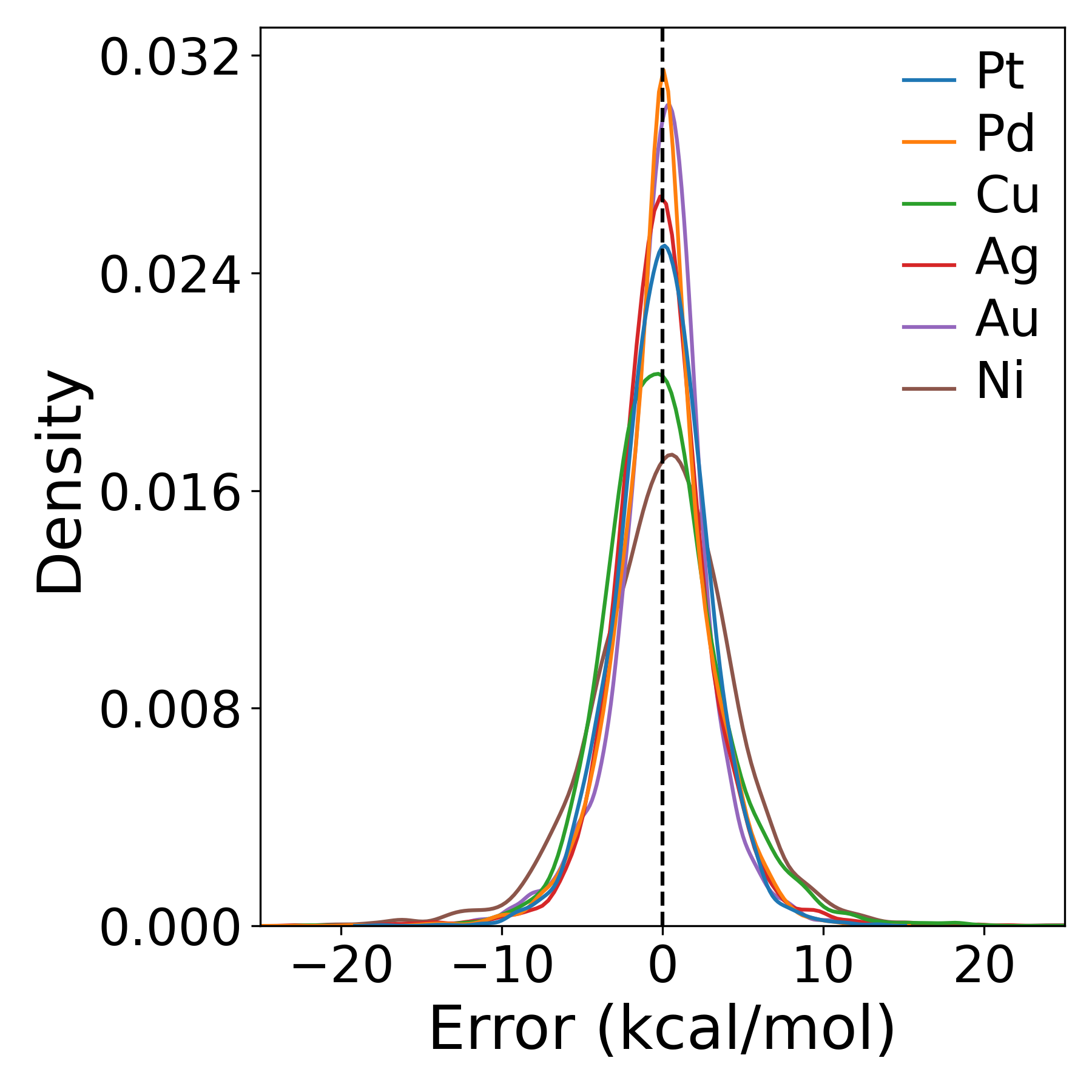}
    \caption{Error distribution per metal from HIP for DB1.}
    \label{fig:err_per_metal}
\end{figure}

\newpage
\section{Supplementary tables}
\begin{table}[h!]
\caption{Performance of the HIP model fit on DB2.}
\centering
\scalebox{1}{
\begin{tabular}{cc}
    \hline
    \textbf{Reaction step} & \Centering MAE (kcal/mol)\\
    \hline
    \textbf{Oxidation} & 3.46\\
    \textbf{Transmetalation} & 0.88\\
    \textbf{REductive elimination} & 3.31\\
    \hline
\end{tabular}}
\label{tab:DB2_results}
\end{table}

\renewcommand{\arraystretch}{1.08} 
\begin{table}[H]
    \caption{\justifying Details of learning curves of KRR direct learning and $\mathrm{\Delta}$-ML models using geometry-based representations and performance curve of cHIP baseline model, showing mean absolute error (MAE) $\pm$ standard deviation.} 
    \label{tab:lc_results}
    \centering
    \scalebox{1}{
    \begin{tabular}{p{2.5cm} l l c c}
    \hline
    \textbf{Method} & \textbf{Representation} & \textbf{Kernel} &\textbf{Training Set Size} & \textbf{MAE (kcal/mol)}\\
    \hline
    \multirow{24}{2.5cm}{Machine learning 7k DFT data} & CM & Laplacian & 250 & 8.03 $\pm$ 0.23\\
       &&& 500 & 5.78 $\pm$ 0.10\\
       &&& 1000& 4.55 $\pm$ 0.06\\
       &&& 2000& 3.74 $\pm$ 0.05\\
       &&& 4000& 3.13 $\pm$ 0.03\\
       &&& 6000& 2.77 $\pm$ 0.02\\
       \cline{2-5}
    &MBDF & Laplacian & 250 & 5.08 $\pm$ 0.17\\
       &&& 500 & 4.45 $\pm$ 0.15\\
       &&& 1000& 3.67 $\pm$ 0.04\\
       &&& 2000& 3.16 $\pm$ 0.05\\
       &&& 4000& 2.71 $\pm$ 0.03\\
       &&& 6000& 2.31 $\pm$ 0.01\\
       \cline{2-5}
    &BoB & Laplacian & 250 & 4.52 $\pm$ 0.10\\
       &&& 500 & 3.93 $\pm$ 0.06\\
       &&& 1000& 3.53 $\pm$ 0.03\\
       &&& 2000& 3.11 $\pm$ 0.03\\
       &&& 4000& 2.68 $\pm$ 0.05\\
       &&& 6000& 2.41 $\pm$ 0.01\\
       \cline{2-5}
    &SLATM & Gaussian & 250 & 3.81 $\pm$ 0.04\\
       &&& 500 & 3.95 $\pm$ 0.03\\
       &&& 1000& 3.23 $\pm$ 0.03\\
       &&& 2000& 2.99 $\pm$ 0.03\\
       &&& 4000& 2.81 $\pm$ 0.02\\
       &&& 6000& 2.69 $\pm$ 0.00\\
       \hline
\end{tabular}}
\end{table}
\begin{table}[H]
    \label{tab:lc_results2}
    \centering
    \scalebox{1}{
    \begin{tabular}{l l l c c}
    \hline
    \textbf{Method} & \textbf{Representation} & \textbf{Kernel} & \textbf{Training set size} & \textbf{MAE (kcal/mol)}\\
    \hline
    \multirow{2}{2.5cm}{cHIP on DB1} &N/A&N/A& 625 & 67.78 $\pm$ 41.89\\
       &&& 1250 & 4.18 $\pm$ 0.27\\
       &&& 2500& 4.11 $\pm$ 0.31\\
       &&& 5000& 4.08 $\pm$ 0.31\\
       &&& 10000& 4.07 $\pm$ 0.31\\
       &&&20000& 4.07 $\pm$ 0.31\\
       \hline
       
    \multirow{2}{2.5cm}{$\Delta_{\mathrm{Hammett}}^{\mathrm{DFT+ML}}$ on DB1} & MBDF & Laplacian
        & 625& 9.40 $\pm$ 0.22\\
       &&& 1250& 2.82 $\pm$ 0.04\\
       &&& 2500& 2.21 $\pm$ 0.02\\
       &&& 5000& 1.62 $\pm$ 0.02\\
       &&&10000& 1.18 $\pm$ 0.01\\
       &&&20000& 0.94 $\pm$ 0.00\\  
       \hline
\end{tabular}}
\end{table}

\begin{table}[h!]
\caption{\justifying Predicted oxidative addition binding free energies, cHIP parameters, and prices of the 30 cheapest newly discovered catalysts. All energy values are in kcal/mol. Prices are relative to that of the cheapest catalyst as per www.sigmaaldrich.com/CA, retrieved on March 8, 2024.}
\begin{filecontents*}{csv/ideal_catalysts_latex.csv}
Ni,4,71,19.95,14.87,34.82,0.86,-62.16,-32.17,1.00
Pd,95,68,-4.53,0.04,-4.49,0.58,-29.44,-32.05,1.50
Pd,83,68,-0.18,0.04,-0.14,0.58,-29.44,-29.52,1.50
Pd,71,68,14.87,0.04,14.91,0.58,-29.44,-20.81,1.50
Pd,83,95,-0.18,-4.53,-4.71,0.58,-29.44,-32.17,1.50
Pd,71,95,14.87,-4.53,10.33,0.58,-29.44,-23.46,1.50
Pd,83,71,-0.18,14.87,14.69,0.58,-29.44,-20.93,1.51
Pd,71,94,14.87,-8.84,6.03,0.58,-29.44,-25.95,1.51
Pd,71,2,14.87,-9.07,5.79,0.58,-29.44,-26.09,1.59
Pd,82,68,-1.34,0.04,-1.30,0.58,-29.44,-30.20,1.61
Pd,82,95,-1.34,-4.53,-5.88,0.58,-29.44,-32.85,1.61
Pd,82,83,-1.34,-0.18,-1.52,0.58,-29.44,-30.33,1.61
Pd,82,71,-1.34,14.87,13.52,0.58,-29.44,-21.61,1.61
Pd,9,68,-0.11,0.04,-0.07,0.58,-29.44,-29.49,1.78
Pd,9,95,-0.11,-4.53,-4.65,0.58,-29.44,-32.14,1.79
Pd,9,83,-0.11,-0.18,-0.29,0.58,-29.44,-29.61,1.79
Pd,9,71,-0.11,14.87,14.75,0.58,-29.44,-20.90,1.79
Pd,9,82,-0.11,-1.34,-1.46,0.58,-29.44,-30.29,1.89
Pd,8,68,-2.99,0.04,-2.95,0.58,-29.44,-31.15,1.91
Pd,8,95,-2.99,-4.53,-7.52,0.58,-29.44,-33.80,1.91
Pd,8,83,-2.99,-0.18,-3.16,0.58,-29.44,-31.28,1.91
Pd,8,71,-2.99,14.87,11.88,0.58,-29.44,-22.56,1.92
Pd,8,82,-2.99,-1.34,-4.33,0.58,-29.44,-31.96,2.02
Pd,9,8,-0.11,-2.99,-3.10,0.58,-29.44,-31.24,2.20
Pd,91,68,0.31,0.04,0.35,0.58,-29.44,-29.24,2.46
Pd,91,95,0.31,-4.53,-4.22,0.58,-29.44,-31.89,2.46
Pd,83,91,-0.18,0.31,0.14,0.58,-29.44,-29.37,2.46
Pd,91,71,0.31,14.87,15.18,0.58,-29.44,-20.65,2.46
Pd,4,68,19.95,0.04,19.99,0.58,-29.44,-17.86,2.49
Pd,4,95,19.95,-4.53,15.42,0.58,-29.44,-20.51,2.49
\end{filecontents*}
\csvreader[
            no head,
            tabular = cccccccccc,
            table head = \toprule \textbf{Metal} & \textbf{Ligand}$\mathbf{_\textit{i}}$ & \textbf{Ligand$\mathbf{_\textit{j}}$} & \textbf{$\mathbf{\sigma_\textit{i}}$} & \textbf{$\mathbf{\sigma_\textit{j}}$} & \textbf{$\mathbf{\sigma_{\textit{ij}}}$} & \textbf{$\mathbf{\rho}$} & \textbf{$\mathbf{\Delta E_{0\textit{m}}}$} & \textbf{$\mathbf{\Delta E_{oxi}}$} & \textbf{Price} \\\midrule,
            table foot = \bottomrule,
        ]{csv/ideal_catalysts_latex.csv}{}{%
            \csvlinetotablerow
        }
\end{table}

\begin{table}[h!]
\centering
\caption{Constants of individual ligands (kcal/mol) from cHIP on DB1.}
\begin{filecontents*}{csv/sigmas_formatted.csv}
Ligand,Sigma,Ligand,Sigma,Ligand,Sigma
0,21.6,37,-7.59,64,-0.41
10,-2.41,38,-7.27,65,-2.48
11,3.76,39,-13.08,66,7.14
12,2.16,3,1.25,67,22.81
13,-2.06,40,-13.76,68,8.08
14,-0.18,41,-8.37,69,1.92
15,8.69,42,-11.82,6,-0.15
16,28.05,43,-5.83,70,0.5
17,-4.68,44,-4.82,71,29.29
18,22.19,45,-11.02,72,-5.32
19,-7.53,46,-6.48,73,10.48
1,19.43,47,-10.55,74,-0.05
20,5.63,48,-10.93,75,1.31
21,0.24,49,-10.2,76,10.67
22,37.28,4,26.94,77,10.75
23,13.43,50,3.61,78,17.07
24,0.74,51,6.86,79,12.08
25,2.86,52,5.4,7,-3.6
26,2.6,53,3.38,80,18.62
27,-3.4,54,3.31,81,3.26
28,17.26,55,3.43,82,-0.18
29,-3.84,56,1.88,83,0.86
2,-3.27,57,0.34,84,20.22
30,14.9,58,2.8,85,-8.59
31,7.43,59,-0.07,86,-13.94
32,-4.77,5,4.51,87,-11.88
33,-9.95,60,-4.22,88,-4.7
34,-4.66,61,1.78,89,-1.94
35,9.56,62,-0.97,8,40.88
36,-12.98,63,-3.6,90,10.95
\end{filecontents*}
\csvreader[
            head to column names,
            tabular = cc|cc|cc,
            table head = \toprule \textbf{Ligand} & $\mathbf{\sigma}$ & \textbf{Ligand} & $\mathbf{\sigma}$ & \textbf{Ligand} & $\mathbf{\sigma}$ \\\midrule,
            table foot = \bottomrule,
        ]{csv/sigmas_formatted.csv}{}{%
            \csvlinetotablerow
        }
\end{table}

\begin{table}[h!]
\centering
\caption{Constants and offsets (kcal/mol) of metals from cHIP on DB1.}
\begin{filecontents*}{csv/metals_formatted.csv}
Field,Rho,x0
Ag,0.82,18.55
Au,0.75,3.13
Cu,0.92,-6.26
Ni,1.00,-63.82
Pd,1.03,-37.33
Pt,0.85,-43.15
\end{filecontents*}
\csvreader[
            head to column names,
            tabular = ccc,
            table head = \toprule \textbf{Metal} & $\mathbf{\rho}$ & $\mathbf{x_0}$ \\\midrule,
            table foot = \bottomrule,
        ]{csv/metals_formatted.csv}{}{%
            \csvlinetotablerow
        }
\end{table}

\clearpage
\section{List of ligands}
\subsection{Ligands from DB1}
The ligands in DB1 were obtained from the publicly-available database published by Meyer \textit{et. al.} [4].
\begin{figure}[h!]
  \centering
  \includegraphics[]{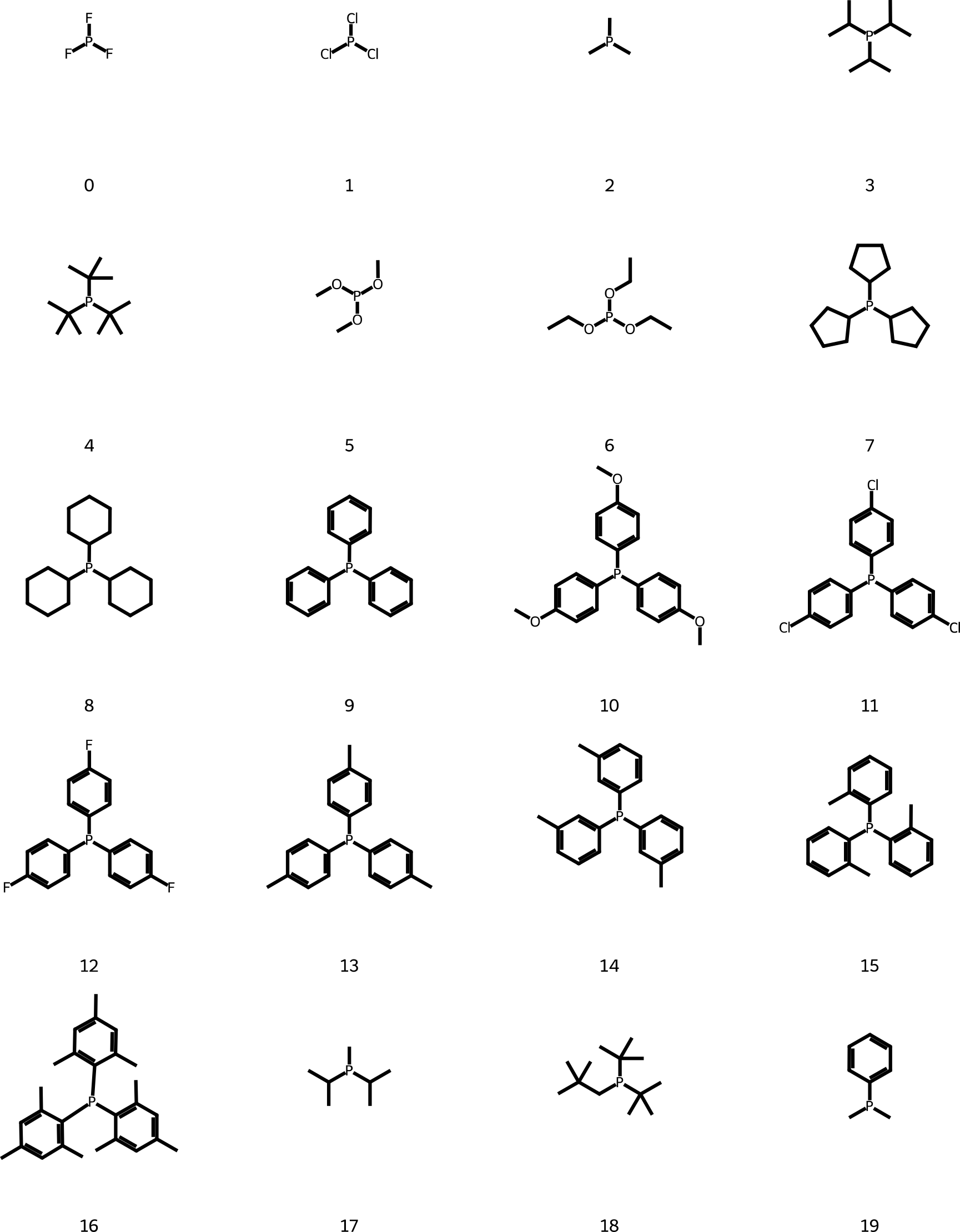}
\end{figure}
\begin{figure}[H]
    \centering   \includegraphics{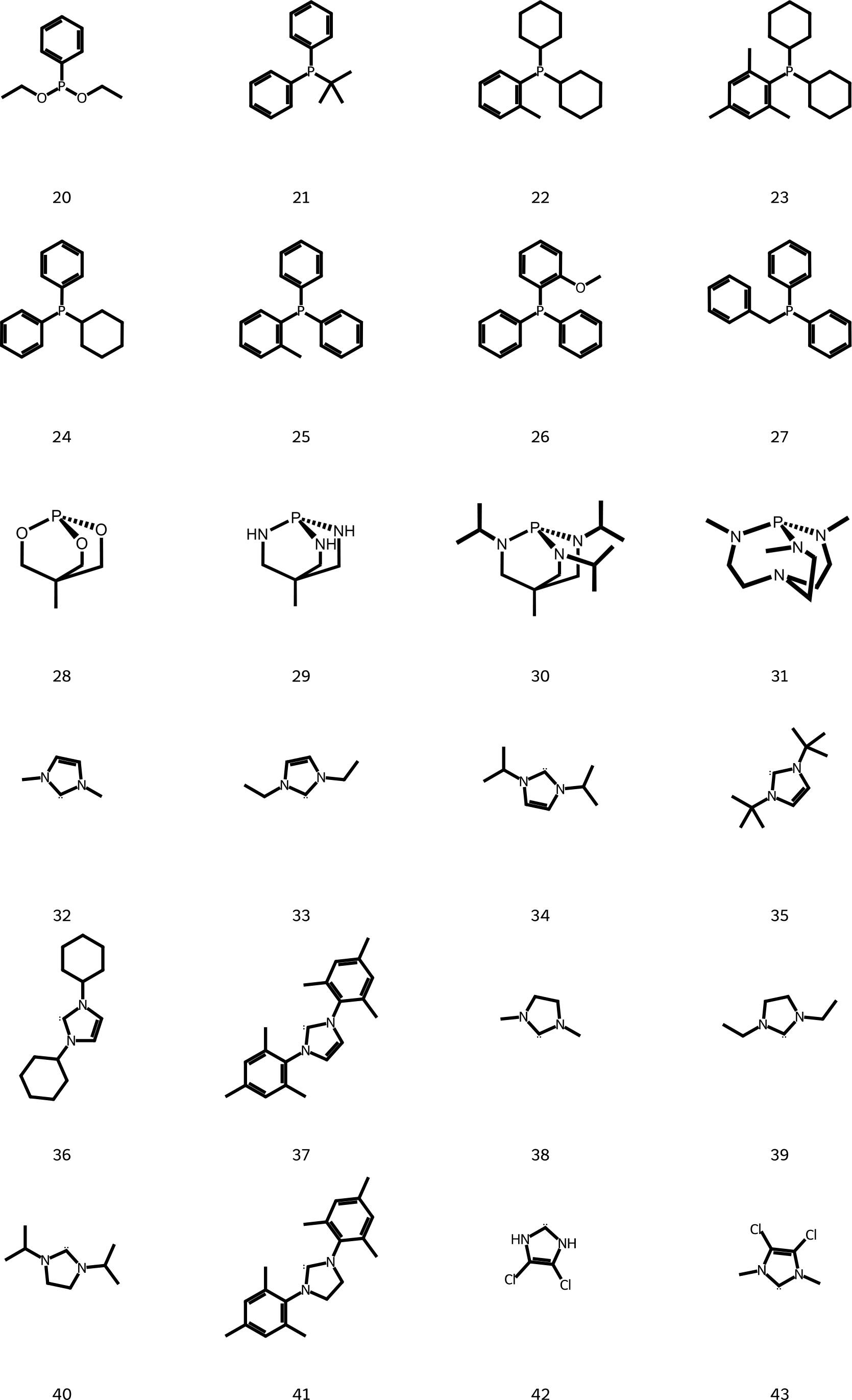}
\end{figure}
\begin{figure}[H]
    \centering
\includegraphics{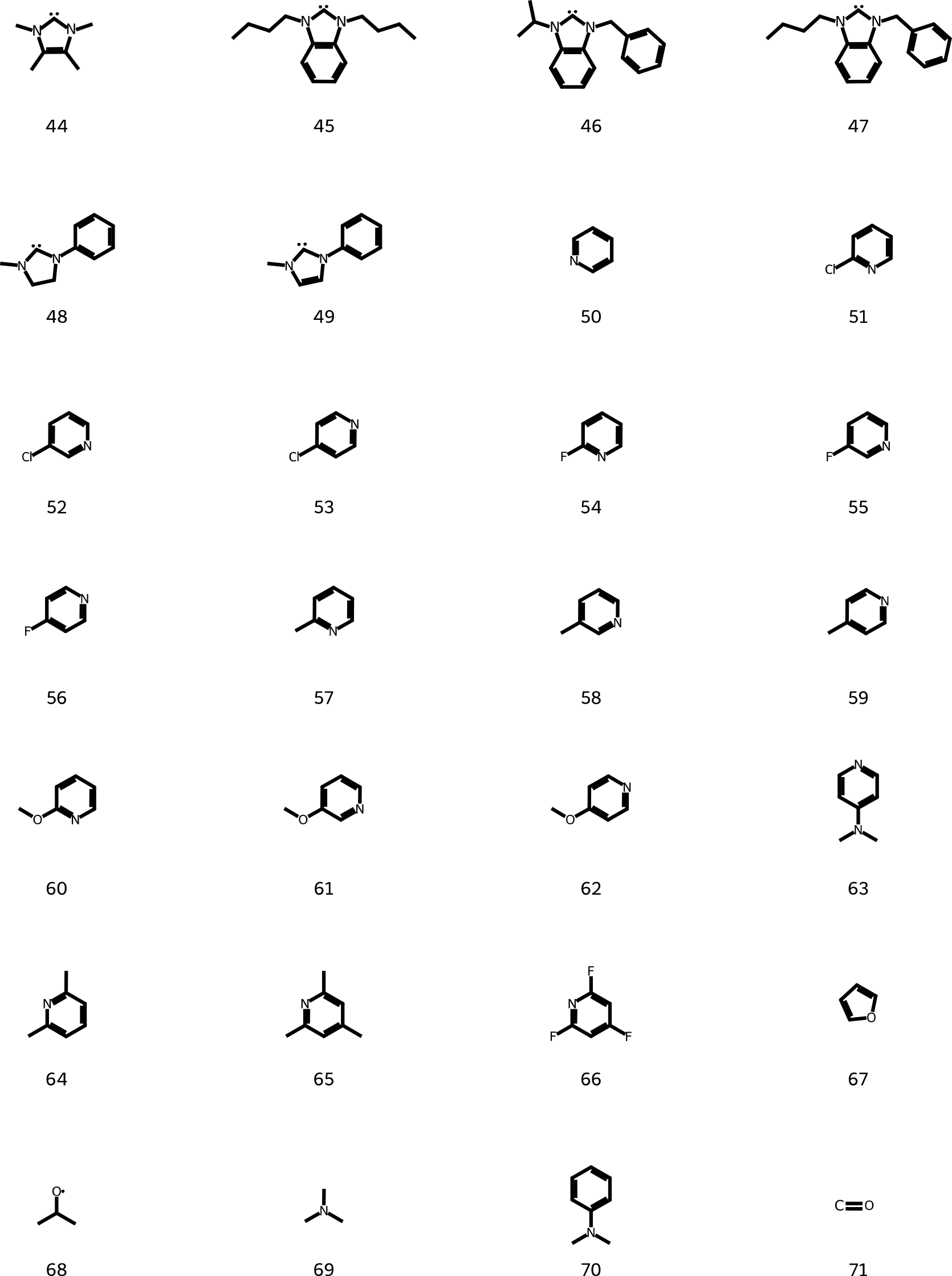}
\end{figure}
\begin{figure}[H]
    \centering\includegraphics{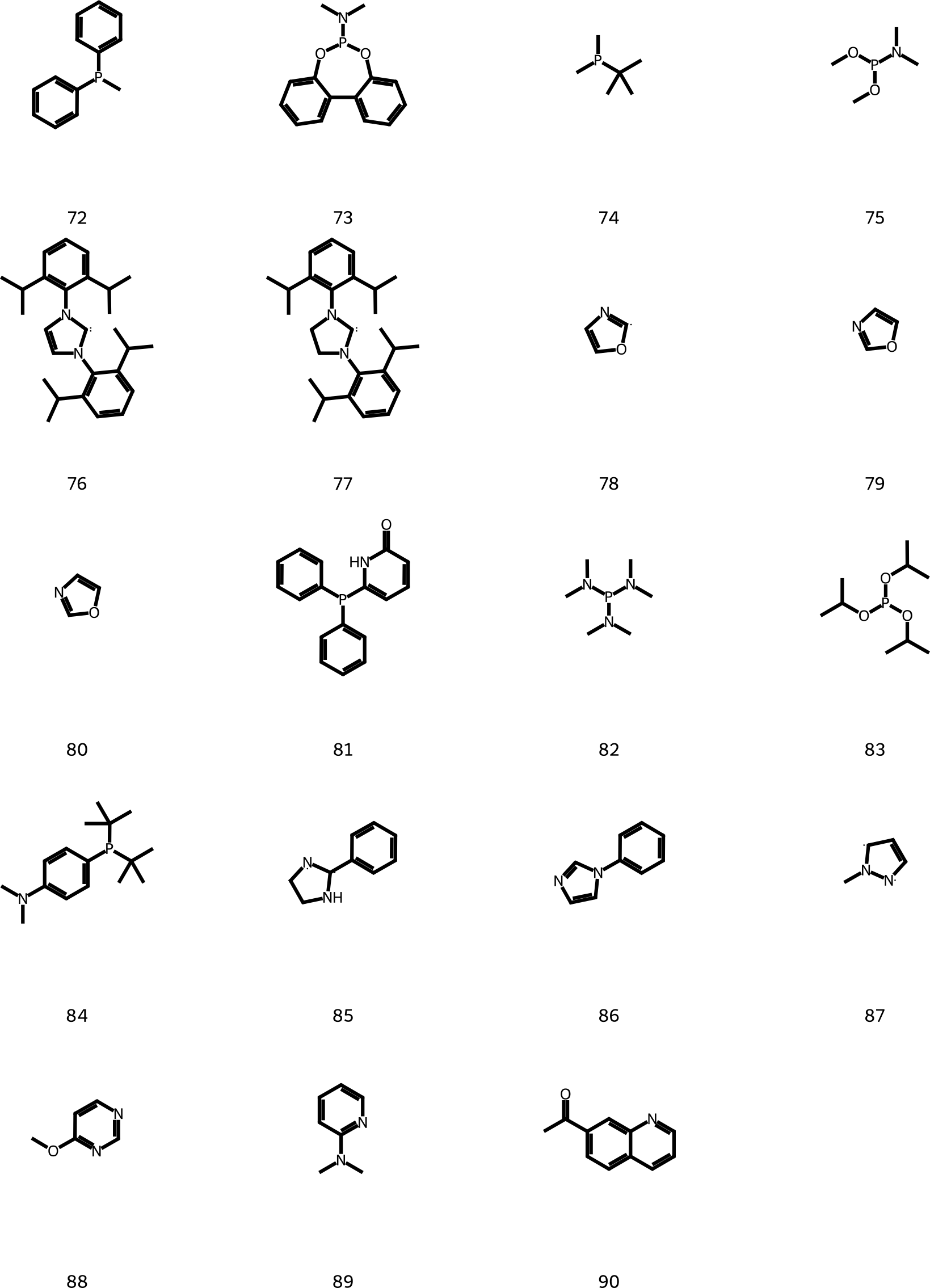}
\end{figure}
\newpage
\subsection{Ligands from DB2}
DB2 [5] also contained Ligands no. 2, 4, 8, 9, 41, 68, 71, 81, 82, 83, and 84 from DB1. The remaining ligands are drawn below. 
\begin{figure}[H]
    \centering
    \includegraphics[scale=0.45]{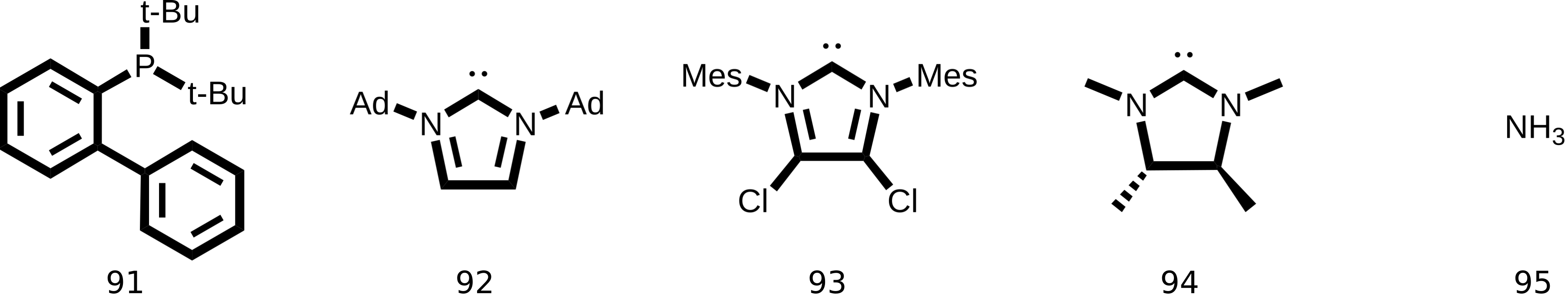}
    \label{fig:DB2}
\end{figure}

\section{References}
$\mathrm{[1]}$ D. Berthelot, Sur le m\'elange des gaz, Compt.
Rendus \textbf{126}, 15 (1898).\\$\mathrm{[2]}$ M. Waldman and A. T. Hagler, New combining
rules for rare gas van der waals parameters,
Journal of computational chemistry \textbf{14}, 1077
(1993).\\$\mathrm{[3]}$  B. Fender and G. Halsey Jr, Second
virial coefficients of argon, krypton, and
argon-krypton mixtures at low temperatures,
The Journal of Chemical Physics \textbf{36}, 1881
(1962).
\\$\mathrm{[4]}$ B. Meyer, B. Sawatlon, S. Heinen, O. A.
von Lilienfeld, and C. Corminboeuf, Machine
learning meets volcano plots: computational
discovery of cross-coupling catalysts, Chemical
science \textbf{9}, 7069 (2018).\\$\mathrm{[5]}$ M. Busch, M. D. Wodrich, and C. Corminboeuf,
A generalized picture of C-C cross-coupling,
ACS Catalysis \textbf{7}, 5643 (2017).